\documentclass[%
 reprint,
 twocolumn,
superscriptaddress,
 amsmath,amssymb,
 aps,
]{revtex4-2}
\usepackage{dsfont}
\usepackage{graphicx}
\usepackage{dcolumn}
\usepackage{bm}
\usepackage{hyperref}
\usepackage{braket}
\usepackage{xcolor}
\usepackage{float}
\usepackage[dvipsnames]{xcolor}
\definecolor{cobalt_blue}{HTML}{1446A0}
\definecolor{mardi_gras}{HTML}{8F2880}

\makeatletter
\def\maketitle{
\@author@finish
\title@column\titleblock@produce
\suppressfloats[t]}
\makeatother

\begin{document}

\newcommand{\TitleName}{Hybrid Path-Transverse Electric Mode Qudit Encoding on an Integrated Photonic Chip}
\title{\TitleName}
\author{Imogen Forbes}
\email{imogen.forbes@bristol.ac.uk}
\affiliation{Quantum Engineering Technology Laboratories, H. H. Wills Physics Laboratory and School of Electrical, Electronic, and Mechanical Engineering, University of Bristol, BS8 1FD, United Kingdom}

\author{Patrick Yard}
\affiliation{Quantum Engineering Technology Laboratories, H. H. Wills Physics Laboratory and School of Electrical, Electronic, and Mechanical Engineering, University of Bristol, BS8 1FD, United Kingdom}

\author{Martin Bielak}
\affiliation{ Department of Optics, Palacký University, 17. listopadu 1192/12, 77900 Olomouc, Czechia}

\author{Molly A. Thomas}
\affiliation{Quantum Engineering Technology Laboratories, H. H. Wills Physics Laboratory and School of Electrical, Electronic, and Mechanical Engineering, University of Bristol, BS8 1FD, United Kingdom}

\author{Matthew S. Jones}
\affiliation{Quantum Engineering Technology Laboratories, H. H. Wills Physics Laboratory and School of Electrical, Electronic, and Mechanical Engineering, University of Bristol, BS8 1FD, United Kingdom}
\affiliation{Quantum Engineering Centre for Doctoral Training, H. H. Wills Physics Laboratory and School of Electrical, Electronic, and Mechanical Engineering, University of Bristol, BS8 1FD, United Kingdom}

\author{Stefano Paesani}
\affiliation{NNF Quantum Computing Programme, Niels Bohr Institute, University of Copenhagen, Copenhagen, Denmark}

\author{Massimo Borghi}
\affiliation{Dipartimento di Fisica, Università di Pavia, Via Agostino Bassi 6, 27100 Pavia, Italy}

\author{Anthony Laing}
\affiliation{Quantum Engineering Technology Laboratories, H. H. Wills Physics Laboratory and School of Electrical, Electronic, and Mechanical Engineering, University of Bristol, BS8 1FD, United Kingdom}

\date{\today}

\begin{abstract}
Hybrid encodings, where multiple degrees of freedom are used to encode quantum information, can increase the size of the Hilbert space with minimal increase to hardware requirements. We show a reprogrammable integrated photonic device, with multimodal components designed to allow for control over the transverse electric modes. We use this device to generate qudit states entangled in the path and transverse electric mode degrees of freedom. We generate and verify a hyperentangled state with a fidelity of $\mathcal{F}_{\text{HE}} = 67.3 \pm 0.2 \%$ and a GHZ$_{4}$-style state with a fidelity of $\mathcal{F}_{\text{GHZ}_{4}} = 85.2 \pm 0.4 \%$. We use our hyperentangled state in a single-copy entanglement distillation protocol, resulting in an average $9.1 \%$ increase in the fidelity of the distilled Bell state for up to a 50\% probability of bit flip error. By utilising degrees of freedom which are readily compatible with integrated photonics, our work highlights how this hybrid encoding demonstrates a first step in using the transverse electric mode to reduce the footprint of integrated quantum photonic experiments.
\end{abstract}

\maketitle

\section{Introduction}

\begin{figure*}[ht!]
\centering
\includegraphics[scale = 0.875]{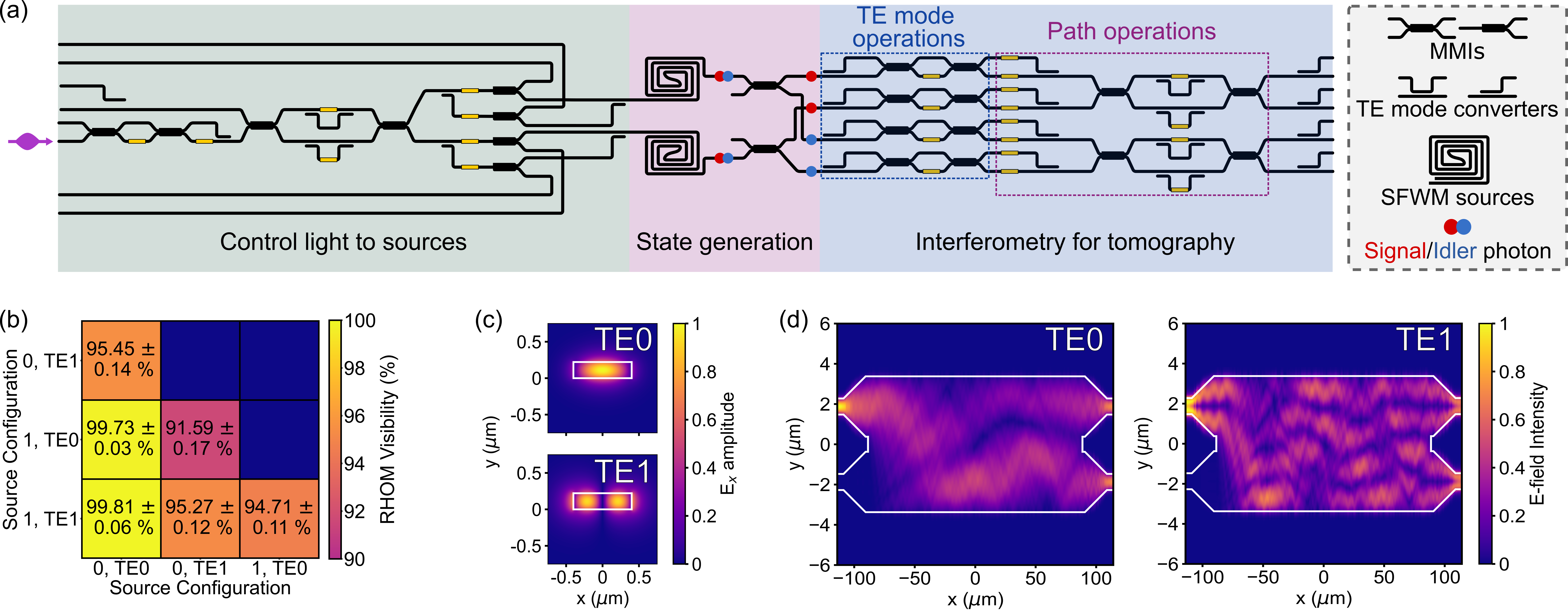}
\caption{\label{fig:experimental_diagram}(a) A schematic of the silicon chip. Pump light is controlled using the interferometry prior to the SFWM sources, where the signal and idler photons are generated. The final section of interferometry allows for tomography to be carried out. (b) RHOM interference visibilities of the waveguide sources when operated to generate different combinations of TE0 and TE1 light. (c) Simulated mode profiles of TE0 and TE1 light in the multimode waveguide. (d) Simulated mode profiles of TE0 and TE1 light in the MMI geometry designed for multimode operation.}
\end{figure*}

Entangled states are a fundamental resource for quantum technologies. Whilst many current applications utilize entanglement between two-level systems, using \textit{qubits}, we can alternatively use a high-dimensional encoding. By using \textit{d}-level quantum systems, or \textit{qudits}, we can increase the dimensionality of the Hilbert space without increasing particle number \cite{kwiat_hyper-entangled_1997}.
Entangled qudit states are beneficial for quantum technologies, such as in quantum communications, where they can provide greater information capacity and the ability to generate secret keys tolerant to higher noise levels in quantum key distribution protocols \cite{zahidy_practical_2024, guo_advances_2019, hu_beating_2018, Nemirovsky-Levy:24, cozzolino_highdqcomms_2019} or in quantum computation, where they offer the potential for fault-tolerant quantum computing with higher error thresholds \cite{PhysRevLett.113.230501, deng_quantum_2017, paesani_highd_2021, wang_qudits_2020}.

Photons are a leading candidate for the generation of entangled states due to their low decoherence and ability to propagate with high fidelities at room temperatures \cite{krastanov_room-temperature_2021}. They also offer a range of different degrees of freedom (DOFs) which can be used to encode information, such as encoding in spatial location \cite{obrien_demonstration_2003}, polarization \cite{Crespi2011}, frequency \cite{Kaneda:19}, time-bins \cite{Marcikic_2004}, or orbital angular momentum \cite{mair_entanglement_2001}.

To make photonic quantum technologies scalable, integrated photonics offers a range of benefits including the ability for devices to be manufactured with pre-existing CMOS foundry processes, with small device footprints, and with better stability than their bulk optical counterparts. These devices also can be networked, with entangled states distributed between chips \cite{Feng_ChiptoChip_2025, Zheng_Multichip_2023}, making them a candidate for future modular quantum architectures. Using a high-dimensional encoding contained solely in the path DOF, where each basis state corresponds to an individual waveguide, has been demonstrated on an integrated platform \cite{chi_high-dimensional_2023}. However, increasing the dimension with a path DOF high-dimensional encoding requires more waveguides and results in an increasingly large device footprint. 

To help mitigate this limitation, we can use additional DOFs to allow for multiple qubits to be encoded per waveguide mode.  Previous experiments have considered combinations of path, polarization, energy-time, time-bin, frequency and orbital angular momentum modes \cite{barreiro_generation_2005, gao_experimental_2010, vallone_hyperentanglement_2009, hu_beating_2018, ciampini_path-polarization_2016, congia_generation_2025}. Many of these DOFs such as time-bin, frequency or polarization encodings require additional experimental components such as frequency converters \cite{wang_2023_quantum}, high-speed modulators \cite{Takesue:09} or path-polarization rotators \cite{shahwar_polarization_2024}, which can be challenging to implement in an integrated photonic platform with the stringent loss requirements of many photonic quantum technologies.

Transverse mode is an attractive secondary DOF in integrated photonics platforms as it can be manipulated fully with passive linear optics. In this work, we present a hybrid encoding using the path and transverse electric (TE) waveguide mode DOFs. In order to achieve control we introduce several new components designed to work simultaneously for several TE modes. Using this device, we generate and verify a GHZ-style state and a hyperentangled state, and demonstrate single-copy entanglement distillation using our hyperentangled state. This encoding offers a potential hardware approach for future modular qudit architectures for quantum information processing. 

\section{Experimental Setup}

A schematic of the silicon chip used in this experiment is shown in Figure \ref{fig:experimental_diagram}(a). A continuous wave pump is incident on two waveguide spirals, each optimised for type-0 phase matched spontaneous four wave mixing (SFWM) in both TE0 and TE1 modes. Prior to the sources, tunable interferometry is used to control the amount of light sent to each source, and which TE mode the light is in. Entanglement is generated by pumping combinations of the spiral sources in different TE modes and post selecting on a single pair of photons being generated \cite{paesani_near-ideal_2020}.

We then use interferometry after the waveguide sources to split the signal and idler photons and to perform unitary operations on the photons to probe the prepared quantum state. To implement unitary operations on the TE mode, we first use asymmetric directional couplers to split the TE modes into separate waveguides followed by a Mach-Zehnder interferometer acting on the two waveguides. A second asymmetric directional coupler is used to recombine the TE modes into a single waveguide. To couple the photons from the chip into fiber, we separate each TE mode to a separate waveguide. The light from the chip is filtered to separate the signal and idler photons and remove pump light. The photons are detected using superconducting nanowire single photon detectors and correlated events are measured via a time tagger. More details on the experimental setup can be found in the Supplemental Material \cite{supp}. 

To allow for on-chip manipulation of the TE mode DOF, the design of the components needs to allow for efficient propagation and conversion between the TE0 and TE1 modes. To this end, we have designed multimode waveguides and multimode interference (MMI) couplers, as well as mode independent waveguide crossers and TE mode converters. Additional details on these components can be found in the Supplemental Material \cite{supp}. Figure 1(c) shows the electric field profiles of the TE modes in the multimode waveguide structure \cite{Lumerical}. MMI couplers that implement a balanced beamsplitter independent of the TE mode of the light are shown in Figure 1(d). 

To characterize the quality of the interference between the four effective single photon sources, we use time-reversed Hong-Ou-Mandel (RHOM) interference  \cite{silverstone_2014}. The visibility of these interference fringes is determined by the overlap of the biphoton state generated by the source. These visibilities are shown in Figure \ref{fig:experimental_diagram}(b) for different combinations of TE0 and TE1 light. Full details of these characterization measurements can be found in the Supplemental Material \cite{supp}.

\section{State Generation}

\begin{figure*}[ht!]
\centering
\includegraphics[scale = 0.875]{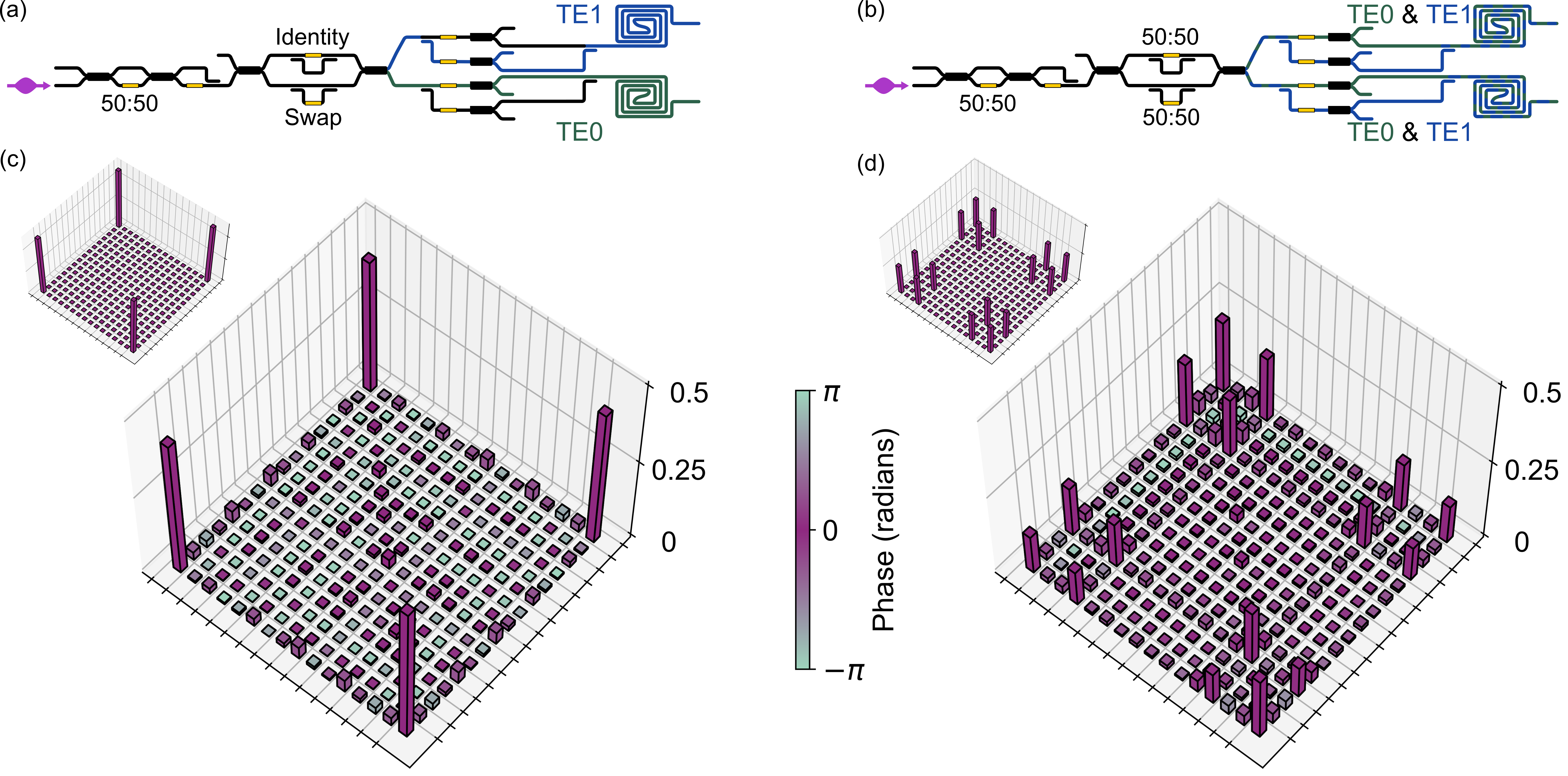}
\caption{\label{fig:resultsHEGHZ} Schematics showing how initial interferometry before the sources is used to pump the sources differently to generate the (a) GHZ$_4$-style state and (b) hyperentangled state. Reconstructed density matrices from the experimentally realized projectors for the (c) GHZ$_4$-style state and (d) hyperentangled states. The insets show the theoretical density matrix for each state.}
\end{figure*}

The reconfigurability of our chip allows a range of different quantum states to be generated using the initial beam splitter network to tune how the sources are pumped. As our chip has four effective single photon sources, it can generate ququart states, which we can interpret as each photon encoding two qubits, one in the path DOF and the other in the TE mode. To demonstrate this, we generate a $\text{GHZ}_4$-style state of the form:
\begin{equation} \label{eq:GHZ}
     \ket{\mathrm{GHZ_4}} =  \frac{1}{2}  \left(\ket{  \mathrm{TE0_0 TE0_1}} \ket{  \mathrm{00}} + \ket{  \mathrm{TE1_0TE1_1} } \ket{  \mathrm{11}} \right),
\end{equation}
and a hyperentangled state of the form:
\begin{equation} \label{eq:HE}
    \ket{\Omega_{HE}} =  \frac{1}{2}   \underbrace{ \left( \ket{  \mathrm{TE0_0 TE0_1}} + \ket{\mathrm{TE1_0TE1_1}} \right) }_{\text{TE Mode}} \otimes   \underbrace{  \left(  \ket{00} + \ket{11} \right) } _{\text{Path}}.
\end{equation}

To characterize the generated states, we reconstruct the density matrix for each state using quantum state tomography. This is carried out by performing projective measurements using the interferometry after the sources on the chip and reconstructing a density matrix using maximum likelihood estimation (MLE) to find the physical state that is most likely to reproduce the measured data \cite{Altepeter2004}. Due to the layout of the phase shifters on our photonic chip, we are unable to access the full set of Pauli operators for the mode qubit to perform complete tomography. To overcome this, we adapt the MLE algorithm to allow the full density matrix to be reconstructed from a restricted set of measurements \cite{Hradil2004}. From our reconstructed density matrix, we can then calculate the fidelity and the entanglement entropy \cite{nielsen_quantum_2010}, where an entanglement entropy of 1 denotes a maximally entangled state. More details on this MLE procedure and entanglement entropy calculation can be found in the Supplemental Material \cite{supp}.

For the $\text{GHZ}_4$-style state, as shown in Figure \ref{fig:resultsHEGHZ}(a), we prepare the state by pumping one of the waveguide sources with TE0 light, and the other with TE1 light. The reconstructed density matrix for this state is shown in Figure \ref{fig:resultsHEGHZ}(c). This state has a fidelity of  ${\mathcal{F}_{\ket{\mathrm{\text{GHZ}_4}}}} = \bra{{\text{GHZ}_4}}\rho_{\text{GHZ}_4}{\ket{{\text{GHZ}_4}}} = 85.2 \pm 0.4 \%$ and an entanglement entropy of $ 0.999 \pm 0.001 $. 

For the hyperentangled state, as shown in Figure \ref{fig:resultsHEGHZ}(b), we prepare the state by pumping both of our waveguide sources with an equal superposition of TE0 and TE1 light. The reconstructed density matrix for this state is shown in Figure \ref{fig:resultsHEGHZ}(d), and the state has a fidelity of $\mathcal{F_\text{HE}} = \bra{\Omega_{HE}}\rho_{HE}{\ket{\Omega_{HE}}} = 67.3 \pm 0.2 \%$. The entanglement entropy is calculated to be $0.998\pm  0.001$, taking the average of the entanglement entropies for the constituent qubits in the state.

\begin{figure*}[ht!]
\centering
\includegraphics[scale = 0.95 ]{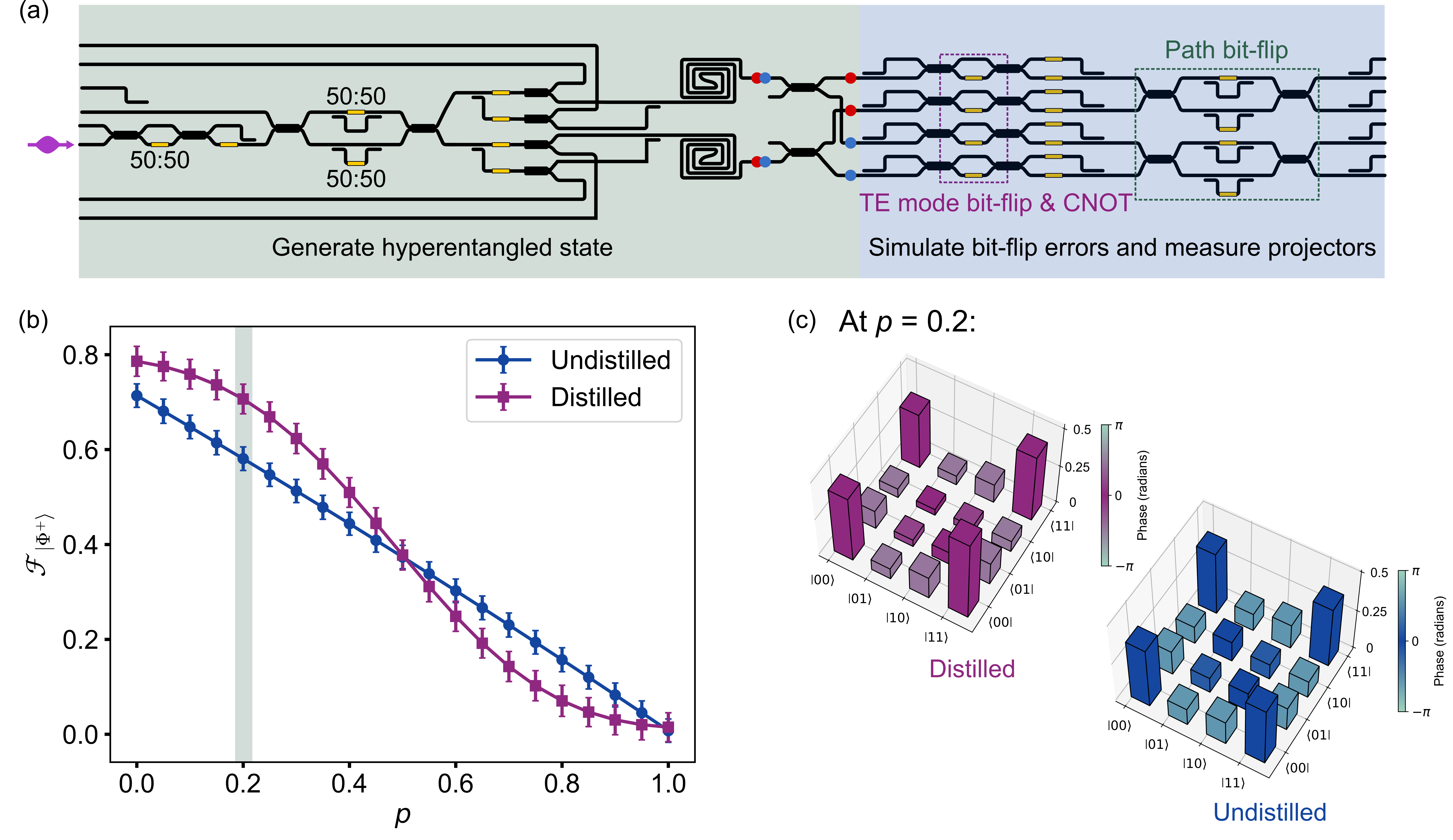}
\caption{\label{fig:entanglementdistillation}(a) Schematic of the device for generating the hyperentangled state. Bit flip errors along with deterministic CNOT operations between the path and mode qubits can be compiled into the single qubit rotations. (b) Plot showing the output fidelities, $\mathcal{F}_{\ket{\Phi^+}}$, of the path Bell state calculated from the stabilizers of the state, in the presence of increasing probability of bit flip error. The blue line shows when the entanglement distillation protocol is not used, and the purple shows when the distillation protocol is used. The green shaded area denotes the points used to reconstruct the density matrices in (c). Errors are calculated from bootstrapping. (c) Reconstructed density matrices for the output Bell state with a 20\% probability of bit flip error, with and without distillation, as highlighted in (b). These density matrices are reconstructed using quantum state tomography with MLE.}
\end{figure*}

A contribution to the infidelity of both states is the imperfect indistinguishability of the photon pair sources, as quantified by the RHOM visibilities shown in Fig. \ref{fig:experimental_diagram}(b) for the different pumping schemes. In particular, we attribute the drop in fidelity from the GHZ$_4$-style state to the hyperentangled state to the number of sources required to prepare each state. For the former we only consider 2 photon pair sources, as each spiral is only pumped with either TE0 or TE1 light, whereas for the latter we are considering 4 photon pair sources, as each spiral is pumped with a superposition of TE0 and TE1 light. As such, the hyperentangled state relies on the indistinguishability of more sources. In addition, there may be crosstalk within the circuit which differs when we alter the phase shifters used for the pumping regimes that generate the different states. With multimodal components, there may also be some spurious mode conversion, both before the sources, affecting the state preparation, or in the beam splitter network used for tomography measurements. Further detail on the components, sources and their characterization can be found in the Supplemental Material \cite{supp}.

\section{Single-Copy Entanglement Distillation}

Entanglement in real systems is subject to noise, which can lead to decoherence and degradation of the entangled states \cite{SCHLOSSHAUER20191}. To mitigate this, state purification methods \cite{yan_entanglementpurification_2023} can be used. For example, recent work has considered purification protocols to target specific noise mechanisms in photonics, namely to decrease the spectral distinguishability of the photons, using a quantum dot source alongside an interferometer \cite{faurby_purifying_2024, hoch_optimal_2025}. At the qubit level, we can consider state purification via entanglement distillation, where multiple copies of a noisy entangled state are used to distill out one lower noise copy \cite{bennett_purification_1996}. This entanglement distillation protocol utilises a CNOT operation between the initial multiple noisy copies of the state. For photonics, this approach poses two challenges: firstly, using multiple copies of the state results in large photon numbers, which are experimentally challenging due to loss \cite{Slussarenko_Photoic_2019}; secondly, any entangling operations can only be implemented probabilistically between photons using only linear optics \cite{pittman_experimental_2003, obrien_demonstration_2003}.
 
An alternative approach is single-copy entanglement distillation \cite{simon_polarization_2002}. Instead of having multiple copies of the noisy entangled state, a single pair of photons can be entangled in multiple degrees of freedom. Entangling operations can then be carried out deterministically. From the initial multiple DOFs in the hyperentangled state, we then measure a higher fidelity Bell state in a single DOF, whilst retaining the same number of quantum particles. This procedure has been demonstrated using bulk optics in \cite{ecker_experimental_2021} for a polarization-time hyperentangled state, and here we demonstrate how this can be implemented with our path and TE mode entangled state on-chip.

To this end, we generate the hyperentangled state, $\ket{\Omega_{HE}},$ as before. Bit flip errors are simulated by including SWAP operations using the Mach-Zehnder interferometers, either for the mode, path, or both DOFs for one of the qubits, as shown in Figure \ref{fig:entanglementdistillation}(a). For the distillation case where a CNOT is applied, the path DOF is treated as the control, and the TE mode as the target. Projective measurements are then taken for the stabilizers, $\mathcal{S}$, of the $\ket{\Phi^+}$ path Bell state, namely $ \mathcal{S} = \{XX, -YY, ZZ\}$ \cite{kok_qip_2010}. These measurements are carried out for the case of no bit flips, a bit flip in path, a bit flip in mode, and both path and mode bit flip errors. We combine the resulting measurement statistics in each of these cases with different weightings to simulate different probabilities of a bit flip error, $p$, on the chip. We estimate the fidelity of the state for different $p$ using the weighted stabilizer measurements as $ \mathcal{F}_{\ket{\Phi^+}} = \frac{1}{4} (1 + \left\langle XX \right\rangle - \left\langle YY \right\rangle + \left\langle ZZ \right\rangle) $. More details on this method can be found in the Supplemental Material \cite{supp}.

As shown in Figure \ref{fig:resultsHEGHZ}(d), our initial hyperentangled resource is noisy, and as a result we see an improvement in the fidelity at $p = 0$ in Figure \ref{fig:entanglementdistillation}(c) - when no additional errors are introduced. The overall protocol is also successful - despite the noise on our original hyperentangled state, in the region from $p = 0$ to $ p = 0.5$, we see an average of a $9.1 \%$ increase in the fidelity with the distillation protocol. This is highlighted through the comparison of the density matrix reconstructions of the Bell states with and without distillation at $p = 0.2$ in Figure \ref{fig:entanglementdistillation}(d). 

\section{Conclusions}
We have demonstrated a hybrid qudit encoding, utilising the TE mode and path DOFs. By using a single, reprogrammable integrated photonic device, we are able to generate a range of high-dimensional states with this encoding, and we have highlighted a hyperentangled state with a fidelity of $\mathcal{F}_{\text{HE}} = 67.3 \pm 0.2 \%$ and a GHZ$_{4}$-style state with a fidelity of $\mathcal{F}_{\text{GHZ}_{4}} = 85.2 \pm 0.4 \%$. The hyperentangled state is then used to carry out on-chip single-copy entanglement distillation, giving an average of a $9.1 \%$  increase in the fidelity for up to 50\% probability of bit flip error. 

This work highlights how the TE mode is a beneficial DOF for implementing hybrid encodings on integrated photonic devices, through the design of multimode components to allow for the manipulation of TE modes on chip. Future devices could prevent the need to convert between TE modes and path by developing multimodal interferometers to reduce the device footprint required for tomography as well as enhancing the multimodal design to allow for higher TE modes to be supported. Development of this work could also consider chip-to-chip protocols to utilize this encoding in modular qudit architectures. Overall, our integrated approach demonstrates the feasibility of hybrid qudit encoding as an alternative to qudit encoding with an increasing number of waveguide modes, with the TE mode being a strong candidate for integrated implementations.

\bigskip
\begin{acknowledgments}
This work was supported by the Engineering and Physical Sciences Research Council (EPSRC) Hub in Quantum Computing and Simulation (EP/T001062/1). M.B. acknowledges the support by Palacký University (projects IGA-PrF-2024-008 and IGA-PrF-2025-010). M.S.J acknowledges financial support from EPSRC Quantum Engineering Centre for Doctoral Training grant EP/SO23607/1. S.P. acknowledges funding from the VILLUM FONDEN research grant No.VIL60743, and support from the NNF Quantum Computing Programme. I.F. would like to thank Edward Deacon and Emilien Lavie for experimental assistance and useful discussions.
\end{acknowledgments}

\bibliography{he_bibliography}

\providecommand{\noopsort}[1]{}\providecommand{\singleletter}[1]{#1}
\begin{thebibliography}{48}%
\makeatletter
\providecommand \@ifxundefined [1]{%
 \@ifx{#1\undefined}
}%
\providecommand \@ifnum [1]{%
 \ifnum #1\expandafter \@firstoftwo
 \else \expandafter \@secondoftwo
 \fi
}%
\providecommand \@ifx [1]{%
 \ifx #1\expandafter \@firstoftwo
 \else \expandafter \@secondoftwo
 \fi
}%
\providecommand \natexlab [1]{#1}%
\providecommand \enquote  [1]{``#1''}%
\providecommand \bibnamefont  [1]{#1}%
\providecommand \bibfnamefont [1]{#1}%
\providecommand \citenamefont [1]{#1}%
\providecommand \href@noop [0]{\@secondoftwo}%
\providecommand \href [0]{\begingroup \@sanitize@url \@href}%
\providecommand \@href[1]{\@@startlink{#1}\@@href}%
\providecommand \@@href[1]{\endgroup#1\@@endlink}%
\providecommand \@sanitize@url [0]{\catcode `\\12\catcode `\$12\catcode `\&12\catcode `\#12\catcode `\^12\catcode `\_12\catcode `\%12\relax}%
\providecommand \@@startlink[1]{}%
\providecommand \@@endlink[0]{}%
\providecommand \url  [0]{\begingroup\@sanitize@url \@url }%
\providecommand \@url [1]{\endgroup\@href {#1}{\urlprefix }}%
\providecommand \urlprefix  [0]{URL }%
\providecommand \Eprint [0]{\href }%
\providecommand \doibase [0]{https://doi.org/}%
\providecommand \selectlanguage [0]{\@gobble}%
\providecommand \bibinfo  [0]{\@secondoftwo}%
\providecommand \bibfield  [0]{\@secondoftwo}%
\providecommand \translation [1]{[#1]}%
\providecommand \BibitemOpen [0]{}%
\providecommand \bibitemStop [0]{}%
\providecommand \bibitemNoStop [0]{.\EOS\space}%
\providecommand \EOS [0]{\spacefactor3000\relax}%
\providecommand \BibitemShut  [1]{\csname bibitem#1\endcsname}%
\let\auto@bib@innerbib\@empty
\bibitem [{\citenamefont {Kwiat}(1997)}]{kwiat_hyper-entangled_1997}%
  \BibitemOpen
  \bibfield  {author} {\bibinfo {author} {\bibfnamefont {P.~G.}\ \bibnamefont {Kwiat}},\ }\bibfield  {title} {\bibinfo {title} {Hyper-entangled states},\ }\href {https://doi.org/10.1080/09500349708231877} {\bibfield  {journal} {\bibinfo  {journal} {Journal of Modern Optics}\ }\textbf {\bibinfo {volume} {44}},\ \bibinfo {pages} {2173} (\bibinfo {year} {1997})}\BibitemShut {NoStop}%
\bibitem [{\citenamefont {Zahidy}\ \emph {et~al.}(2024)\citenamefont {Zahidy}, \citenamefont {Ribezzo}, \citenamefont {De~Lazzari}, \citenamefont {Vagniluca}, \citenamefont {Biagi}, \citenamefont {Müller}, \citenamefont {Occhipinti}, \citenamefont {Oxenløwe}, \citenamefont {Galili}, \citenamefont {Hayashi}, \citenamefont {Cassioli}, \citenamefont {Mecozzi}, \citenamefont {Antonelli}, \citenamefont {Zavatta},\ and\ \citenamefont {Bacco}}]{zahidy_practical_2024}%
  \BibitemOpen
  \bibfield  {author} {\bibinfo {author} {\bibfnamefont {M.}~\bibnamefont {Zahidy}}, \bibinfo {author} {\bibfnamefont {D.}~\bibnamefont {Ribezzo}}, \bibinfo {author} {\bibfnamefont {C.}~\bibnamefont {De~Lazzari}}, \bibinfo {author} {\bibfnamefont {I.}~\bibnamefont {Vagniluca}}, \bibinfo {author} {\bibfnamefont {N.}~\bibnamefont {Biagi}}, \bibinfo {author} {\bibfnamefont {R.}~\bibnamefont {Müller}}, \bibinfo {author} {\bibfnamefont {T.}~\bibnamefont {Occhipinti}}, \bibinfo {author} {\bibfnamefont {L.~K.}\ \bibnamefont {Oxenløwe}}, \bibinfo {author} {\bibfnamefont {M.}~\bibnamefont {Galili}}, \bibinfo {author} {\bibfnamefont {T.}~\bibnamefont {Hayashi}}, \bibinfo {author} {\bibfnamefont {D.}~\bibnamefont {Cassioli}}, \bibinfo {author} {\bibfnamefont {A.}~\bibnamefont {Mecozzi}}, \bibinfo {author} {\bibfnamefont {C.}~\bibnamefont {Antonelli}}, \bibinfo {author} {\bibfnamefont {A.}~\bibnamefont {Zavatta}},\ and\ \bibinfo {author} {\bibfnamefont {D.}~\bibnamefont {Bacco}},\ }\bibfield  {title} {\bibinfo {title}
  {Practical high-dimensional quantum key distribution protocol over deployed multicore fiber},\ }\href {https://doi.org/10.1038/s41467-024-45876-x} {\bibfield  {journal} {\bibinfo  {journal} {Nature Communications}\ }\textbf {\bibinfo {volume} {15}},\ \bibinfo {pages} {1651} (\bibinfo {year} {2024})}\BibitemShut {NoStop}%
\bibitem [{\citenamefont {Guo}\ \emph {et~al.}(2019)\citenamefont {Guo}, \citenamefont {Liu}, \citenamefont {Li},\ and\ \citenamefont {Guo}}]{guo_advances_2019}%
  \BibitemOpen
  \bibfield  {author} {\bibinfo {author} {\bibfnamefont {Y.}~\bibnamefont {Guo}}, \bibinfo {author} {\bibfnamefont {B.-H.}\ \bibnamefont {Liu}}, \bibinfo {author} {\bibfnamefont {C.-F.}\ \bibnamefont {Li}},\ and\ \bibinfo {author} {\bibfnamefont {G.-C.}\ \bibnamefont {Guo}},\ }\bibfield  {title} {\bibinfo {title} {Advances in {Quantum} {Dense} {Coding}},\ }\href {https://doi.org/10.1002/qute.201900011} {\bibfield  {journal} {\bibinfo  {journal} {Advanced Quantum Technologies}\ }\textbf {\bibinfo {volume} {2}},\ \bibinfo {pages} {1900011} (\bibinfo {year} {2019})}\BibitemShut {NoStop}%
\bibitem [{\citenamefont {Hu}\ \emph {et~al.}(2018)\citenamefont {Hu}, \citenamefont {Guo}, \citenamefont {Liu}, \citenamefont {Huang}, \citenamefont {Li},\ and\ \citenamefont {Guo}}]{hu_beating_2018}%
  \BibitemOpen
  \bibfield  {author} {\bibinfo {author} {\bibfnamefont {X.-M.}\ \bibnamefont {Hu}}, \bibinfo {author} {\bibfnamefont {Y.}~\bibnamefont {Guo}}, \bibinfo {author} {\bibfnamefont {B.-H.}\ \bibnamefont {Liu}}, \bibinfo {author} {\bibfnamefont {Y.-F.}\ \bibnamefont {Huang}}, \bibinfo {author} {\bibfnamefont {C.-F.}\ \bibnamefont {Li}},\ and\ \bibinfo {author} {\bibfnamefont {G.-C.}\ \bibnamefont {Guo}},\ }\bibfield  {title} {\bibinfo {title} {Beating the channel capacity limit for superdense coding with entangled ququarts},\ }\href {https://doi.org/10.1126/sciadv.aat9304} {\bibfield  {journal} {\bibinfo  {journal} {Sci. Adv.}\ }\textbf {\bibinfo {volume} {4}},\ \bibinfo {pages} {eaat9304} (\bibinfo {year} {2018})}\BibitemShut {NoStop}%
\bibitem [{\citenamefont {Nemirovsky-Levy}\ \emph {et~al.}(2024)\citenamefont {Nemirovsky-Levy}, \citenamefont {Pereg},\ and\ \citenamefont {Segev}}]{Nemirovsky-Levy:24}%
  \BibitemOpen
  \bibfield  {author} {\bibinfo {author} {\bibfnamefont {L.}~\bibnamefont {Nemirovsky-Levy}}, \bibinfo {author} {\bibfnamefont {U.}~\bibnamefont {Pereg}},\ and\ \bibinfo {author} {\bibfnamefont {M.}~\bibnamefont {Segev}},\ }\bibfield  {title} {\bibinfo {title} {Increasing quantum communication rates using hyperentangled photonic states},\ }\href@noop {} {\bibfield  {journal} {\bibinfo  {journal} {Optica Quantum}\ }\textbf {\bibinfo {volume} {2}},\ \bibinfo {pages} {165} (\bibinfo {year} {2024})}\BibitemShut {NoStop}%
\bibitem [{\citenamefont {Cozzolino}\ \emph {et~al.}(2019)\citenamefont {Cozzolino}, \citenamefont {Da~Lio}, \citenamefont {Bacco},\ and\ \citenamefont {Oxenløwe}}]{cozzolino_highdqcomms_2019}%
  \BibitemOpen
  \bibfield  {author} {\bibinfo {author} {\bibfnamefont {D.}~\bibnamefont {Cozzolino}}, \bibinfo {author} {\bibfnamefont {B.}~\bibnamefont {Da~Lio}}, \bibinfo {author} {\bibfnamefont {D.}~\bibnamefont {Bacco}},\ and\ \bibinfo {author} {\bibfnamefont {L.~K.}\ \bibnamefont {Oxenløwe}},\ }\bibfield  {title} {\bibinfo {title} {High-dimensional quantum communication: Benefits, progress, and future challenges},\ }\href {https://doi.org/https://doi.org/10.1002/qute.201900038} {\bibfield  {journal} {\bibinfo  {journal} {Advanced Quantum Technologies}\ }\textbf {\bibinfo {volume} {2}},\ \bibinfo {pages} {1900038} (\bibinfo {year} {2019})}\BibitemShut {NoStop}%
\bibitem [{\citenamefont {Campbell}(2014)}]{PhysRevLett.113.230501}%
  \BibitemOpen
  \bibfield  {author} {\bibinfo {author} {\bibfnamefont {E.~T.}\ \bibnamefont {Campbell}},\ }\bibfield  {title} {\bibinfo {title} {Enhanced fault-tolerant quantum computing in $d$-level systems},\ }\href {https://doi.org/10.1103/PhysRevLett.113.230501} {\bibfield  {journal} {\bibinfo  {journal} {Phys. Rev. Lett.}\ }\textbf {\bibinfo {volume} {113}},\ \bibinfo {pages} {230501} (\bibinfo {year} {2014})}\BibitemShut {NoStop}%
\bibitem [{\citenamefont {Deng}\ \emph {et~al.}(2017)\citenamefont {Deng}, \citenamefont {Ren},\ and\ \citenamefont {Li}}]{deng_quantum_2017}%
  \BibitemOpen
  \bibfield  {author} {\bibinfo {author} {\bibfnamefont {F.-G.}\ \bibnamefont {Deng}}, \bibinfo {author} {\bibfnamefont {B.-C.}\ \bibnamefont {Ren}},\ and\ \bibinfo {author} {\bibfnamefont {X.-H.}\ \bibnamefont {Li}},\ }\bibfield  {title} {\bibinfo {title} {Quantum hyperentanglement and its applications in quantum information processing},\ }\href {https://doi.org/10.1016/j.scib.2016.11.007} {\bibfield  {journal} {\bibinfo  {journal} {Science Bulletin}\ }\textbf {\bibinfo {volume} {62}},\ \bibinfo {pages} {46} (\bibinfo {year} {2017})}\BibitemShut {NoStop}%
\bibitem [{\citenamefont {Paesani}\ \emph {et~al.}(2021)\citenamefont {Paesani}, \citenamefont {Bulmer}, \citenamefont {Jones}, \citenamefont {Santagati},\ and\ \citenamefont {Laing}}]{paesani_highd_2021}%
  \BibitemOpen
  \bibfield  {author} {\bibinfo {author} {\bibfnamefont {S.}~\bibnamefont {Paesani}}, \bibinfo {author} {\bibfnamefont {J.~F.~F.}\ \bibnamefont {Bulmer}}, \bibinfo {author} {\bibfnamefont {A.~E.}\ \bibnamefont {Jones}}, \bibinfo {author} {\bibfnamefont {R.}~\bibnamefont {Santagati}},\ and\ \bibinfo {author} {\bibfnamefont {A.}~\bibnamefont {Laing}},\ }\bibfield  {title} {\bibinfo {title} {Scheme for universal high-dimensional quantum computation with linear optics},\ }\href {https://doi.org/10.1103/PhysRevLett.126.230504} {\bibfield  {journal} {\bibinfo  {journal} {Phys. Rev. Lett.}\ }\textbf {\bibinfo {volume} {126}},\ \bibinfo {pages} {230504} (\bibinfo {year} {2021})}\BibitemShut {NoStop}%
\bibitem [{\citenamefont {Wang}\ \emph {et~al.}(2020)\citenamefont {Wang}, \citenamefont {Hu}, \citenamefont {Sanders},\ and\ \citenamefont {Kais}}]{wang_qudits_2020}%
  \BibitemOpen
  \bibfield  {author} {\bibinfo {author} {\bibfnamefont {Y.}~\bibnamefont {Wang}}, \bibinfo {author} {\bibfnamefont {Z.}~\bibnamefont {Hu}}, \bibinfo {author} {\bibfnamefont {B.~C.}\ \bibnamefont {Sanders}},\ and\ \bibinfo {author} {\bibfnamefont {S.}~\bibnamefont {Kais}},\ }\bibfield  {title} {\bibinfo {title} {Qudits and high-dimensional quantum computing},\ }\bibfield  {journal} {\bibinfo  {journal} {Frontiers in Physics}\ }\textbf {\bibinfo {volume} {8}},\ \href {https://doi.org/10.3389/fphy.2020.589504} {10.3389/fphy.2020.589504} (\bibinfo {year} {2020})\BibitemShut {NoStop}%
\bibitem [{\citenamefont {Krastanov}\ \emph {et~al.}(2021)\citenamefont {Krastanov}, \citenamefont {Heuck}, \citenamefont {Shapiro}, \citenamefont {Narang}, \citenamefont {Englund},\ and\ \citenamefont {Jacobs}}]{krastanov_room-temperature_2021}%
  \BibitemOpen
  \bibfield  {author} {\bibinfo {author} {\bibfnamefont {S.}~\bibnamefont {Krastanov}}, \bibinfo {author} {\bibfnamefont {M.}~\bibnamefont {Heuck}}, \bibinfo {author} {\bibfnamefont {J.~H.}\ \bibnamefont {Shapiro}}, \bibinfo {author} {\bibfnamefont {P.}~\bibnamefont {Narang}}, \bibinfo {author} {\bibfnamefont {D.~R.}\ \bibnamefont {Englund}},\ and\ \bibinfo {author} {\bibfnamefont {K.}~\bibnamefont {Jacobs}},\ }\bibfield  {title} {\bibinfo {title} {Room-temperature photonic logical qubits via second-order nonlinearities},\ }\href {https://doi.org/10.1038/s41467-020-20417-4} {\bibfield  {journal} {\bibinfo  {journal} {Nat Commun}\ }\textbf {\bibinfo {volume} {12}},\ \bibinfo {pages} {191} (\bibinfo {year} {2021})}\BibitemShut {NoStop}%
\bibitem [{\citenamefont {O'Brien}\ \emph {et~al.}(2003)\citenamefont {O'Brien}, \citenamefont {Pryde}, \citenamefont {White}, \citenamefont {Ralph},\ and\ \citenamefont {Branning}}]{obrien_demonstration_2003}%
  \BibitemOpen
  \bibfield  {author} {\bibinfo {author} {\bibfnamefont {J.~L.}\ \bibnamefont {O'Brien}}, \bibinfo {author} {\bibfnamefont {G.~J.}\ \bibnamefont {Pryde}}, \bibinfo {author} {\bibfnamefont {A.~G.}\ \bibnamefont {White}}, \bibinfo {author} {\bibfnamefont {T.~C.}\ \bibnamefont {Ralph}},\ and\ \bibinfo {author} {\bibfnamefont {D.}~\bibnamefont {Branning}},\ }\bibfield  {title} {\bibinfo {title} {Demonstration of an all-optical quantum controlled-{NOT} gate},\ }\href {https://doi.org/10.1038/nature02054} {\bibfield  {journal} {\bibinfo  {journal} {Nature}\ }\textbf {\bibinfo {volume} {426}},\ \bibinfo {pages} {264} (\bibinfo {year} {2003})}\BibitemShut {NoStop}%
\bibitem [{\citenamefont {Crespi}\ \emph {et~al.}(2011)\citenamefont {Crespi}, \citenamefont {Ramponi}, \citenamefont {Osellame}, \citenamefont {Sansoni}, \citenamefont {Bongioanni}, \citenamefont {Sciarrino}, \citenamefont {Vallone},\ and\ \citenamefont {Mataloni}}]{Crespi2011}%
  \BibitemOpen
  \bibfield  {author} {\bibinfo {author} {\bibfnamefont {A.}~\bibnamefont {Crespi}}, \bibinfo {author} {\bibfnamefont {R.}~\bibnamefont {Ramponi}}, \bibinfo {author} {\bibfnamefont {R.}~\bibnamefont {Osellame}}, \bibinfo {author} {\bibfnamefont {L.}~\bibnamefont {Sansoni}}, \bibinfo {author} {\bibfnamefont {I.}~\bibnamefont {Bongioanni}}, \bibinfo {author} {\bibfnamefont {F.}~\bibnamefont {Sciarrino}}, \bibinfo {author} {\bibfnamefont {G.}~\bibnamefont {Vallone}},\ and\ \bibinfo {author} {\bibfnamefont {P.}~\bibnamefont {Mataloni}},\ }\bibfield  {title} {\bibinfo {title} {Integrated photonic quantum gates for polarization qubits},\ }\href {http://dx.doi.org/10.1038/ncomms1570} {\bibfield  {journal} {\bibinfo  {journal} {Nat. Commun.}\ }\textbf {\bibinfo {volume} {2}},\ \bibinfo {pages} {566} (\bibinfo {year} {2011})}\BibitemShut {NoStop}%
\bibitem [{\citenamefont {Kaneda}\ \emph {et~al.}(2019)\citenamefont {Kaneda}, \citenamefont {Suzuki}, \citenamefont {Shimizu},\ and\ \citenamefont {Edamatsu}}]{Kaneda:19}%
  \BibitemOpen
  \bibfield  {author} {\bibinfo {author} {\bibfnamefont {F.}~\bibnamefont {Kaneda}}, \bibinfo {author} {\bibfnamefont {H.}~\bibnamefont {Suzuki}}, \bibinfo {author} {\bibfnamefont {R.}~\bibnamefont {Shimizu}},\ and\ \bibinfo {author} {\bibfnamefont {K.}~\bibnamefont {Edamatsu}},\ }\bibfield  {title} {\bibinfo {title} {Direct generation of frequency-bin entangled photons via two-period quasi-phase-matched parametric downconversion},\ }\href {https://doi.org/10.1364/OE.27.001416} {\bibfield  {journal} {\bibinfo  {journal} {Opt. Express}\ }\textbf {\bibinfo {volume} {27}},\ \bibinfo {pages} {1416} (\bibinfo {year} {2019})}\BibitemShut {NoStop}%
\bibitem [{\citenamefont {Marcikic}\ \emph {et~al.}(2004)\citenamefont {Marcikic}, \citenamefont {de~Riedmatten}, \citenamefont {Tittel}, \citenamefont {Zbinden}, \citenamefont {Legr\'e},\ and\ \citenamefont {Gisin}}]{Marcikic_2004}%
  \BibitemOpen
  \bibfield  {author} {\bibinfo {author} {\bibfnamefont {I.}~\bibnamefont {Marcikic}}, \bibinfo {author} {\bibfnamefont {H.}~\bibnamefont {de~Riedmatten}}, \bibinfo {author} {\bibfnamefont {W.}~\bibnamefont {Tittel}}, \bibinfo {author} {\bibfnamefont {H.}~\bibnamefont {Zbinden}}, \bibinfo {author} {\bibfnamefont {M.}~\bibnamefont {Legr\'e}},\ and\ \bibinfo {author} {\bibfnamefont {N.}~\bibnamefont {Gisin}},\ }\bibfield  {title} {\bibinfo {title} {Distribution of time-bin entangled qubits over 50 km of optical fiber},\ }\href {https://doi.org/10.1103/PhysRevLett.93.180502} {\bibfield  {journal} {\bibinfo  {journal} {Phys. Rev. Lett.}\ }\textbf {\bibinfo {volume} {93}},\ \bibinfo {pages} {180502} (\bibinfo {year} {2004})}\BibitemShut {NoStop}%
\bibitem [{\citenamefont {Mair}\ \emph {et~al.}(2001)\citenamefont {Mair}, \citenamefont {Vaziri}, \citenamefont {Weihs},\ and\ \citenamefont {Zeilinger}}]{mair_entanglement_2001}%
  \BibitemOpen
  \bibfield  {author} {\bibinfo {author} {\bibfnamefont {A.}~\bibnamefont {Mair}}, \bibinfo {author} {\bibfnamefont {A.}~\bibnamefont {Vaziri}}, \bibinfo {author} {\bibfnamefont {G.}~\bibnamefont {Weihs}},\ and\ \bibinfo {author} {\bibfnamefont {A.}~\bibnamefont {Zeilinger}},\ }\bibfield  {title} {\bibinfo {title} {Entanglement of the orbital angular momentum states of photons},\ }\href {https://doi.org/10.1038/35085529} {\bibfield  {journal} {\bibinfo  {journal} {Nature}\ }\textbf {\bibinfo {volume} {412}},\ \bibinfo {pages} {313} (\bibinfo {year} {2001})}\BibitemShut {NoStop}%
\bibitem [{\citenamefont {Feng}\ \emph {et~al.}(2025)\citenamefont {Feng}, \citenamefont {Zhang}, \citenamefont {Liu}, \citenamefont {Cheng}, \citenamefont {Song}, \citenamefont {Ding}, \citenamefont {Dai}, \citenamefont {Guo}, \citenamefont {Guo},\ and\ \citenamefont {Ren}}]{Feng_ChiptoChip_2025}%
  \BibitemOpen
  \bibfield  {author} {\bibinfo {author} {\bibfnamefont {L.-T.}\ \bibnamefont {Feng}}, \bibinfo {author} {\bibfnamefont {M.}~\bibnamefont {Zhang}}, \bibinfo {author} {\bibfnamefont {D.}~\bibnamefont {Liu}}, \bibinfo {author} {\bibfnamefont {Y.-J.}\ \bibnamefont {Cheng}}, \bibinfo {author} {\bibfnamefont {X.-Y.}\ \bibnamefont {Song}}, \bibinfo {author} {\bibfnamefont {Y.-Y.}\ \bibnamefont {Ding}}, \bibinfo {author} {\bibfnamefont {D.-X.}\ \bibnamefont {Dai}}, \bibinfo {author} {\bibfnamefont {G.-P.}\ \bibnamefont {Guo}}, \bibinfo {author} {\bibfnamefont {G.-C.}\ \bibnamefont {Guo}},\ and\ \bibinfo {author} {\bibfnamefont {X.-F.}\ \bibnamefont {Ren}},\ }\bibfield  {title} {\bibinfo {title} {Chip-to-chip quantum photonic controlled-not gate teleportation},\ }\href {https://doi.org/10.1103/d53g-v8q6} {\bibfield  {journal} {\bibinfo  {journal} {Phys. Rev. Lett.}\ }\textbf {\bibinfo {volume} {135}},\ \bibinfo {pages} {020802} (\bibinfo {year} {2025})}\BibitemShut {NoStop}%
\bibitem [{\citenamefont {Zheng}\ \emph {et~al.}(2023)\citenamefont {Zheng}, \citenamefont {Zhai}, \citenamefont {Liu}, \citenamefont {Mao}, \citenamefont {Chen}, \citenamefont {Dai}, \citenamefont {Huang}, \citenamefont {Bao}, \citenamefont {Fu}, \citenamefont {Tong}, \citenamefont {Zhou}, \citenamefont {Yang}, \citenamefont {Tang}, \citenamefont {Li}, \citenamefont {Li}, \citenamefont {Gong}, \citenamefont {Tsang}, \citenamefont {Dai},\ and\ \citenamefont {Wang}}]{Zheng_Multichip_2023}%
  \BibitemOpen
  \bibfield  {author} {\bibinfo {author} {\bibfnamefont {Y.}~\bibnamefont {Zheng}}, \bibinfo {author} {\bibfnamefont {C.}~\bibnamefont {Zhai}}, \bibinfo {author} {\bibfnamefont {D.}~\bibnamefont {Liu}}, \bibinfo {author} {\bibfnamefont {J.}~\bibnamefont {Mao}}, \bibinfo {author} {\bibfnamefont {X.}~\bibnamefont {Chen}}, \bibinfo {author} {\bibfnamefont {T.}~\bibnamefont {Dai}}, \bibinfo {author} {\bibfnamefont {J.}~\bibnamefont {Huang}}, \bibinfo {author} {\bibfnamefont {J.}~\bibnamefont {Bao}}, \bibinfo {author} {\bibfnamefont {Z.}~\bibnamefont {Fu}}, \bibinfo {author} {\bibfnamefont {Y.}~\bibnamefont {Tong}}, \bibinfo {author} {\bibfnamefont {X.}~\bibnamefont {Zhou}}, \bibinfo {author} {\bibfnamefont {Y.}~\bibnamefont {Yang}}, \bibinfo {author} {\bibfnamefont {B.}~\bibnamefont {Tang}}, \bibinfo {author} {\bibfnamefont {Z.}~\bibnamefont {Li}}, \bibinfo {author} {\bibfnamefont {Y.}~\bibnamefont {Li}}, \bibinfo {author} {\bibfnamefont {Q.}~\bibnamefont {Gong}}, \bibinfo {author} {\bibfnamefont {H.~K.}\
  \bibnamefont {Tsang}}, \bibinfo {author} {\bibfnamefont {D.}~\bibnamefont {Dai}},\ and\ \bibinfo {author} {\bibfnamefont {J.}~\bibnamefont {Wang}},\ }\bibfield  {title} {\bibinfo {title} {Multichip multidimensional quantum networks with entanglement retrievability},\ }\href {https://doi.org/10.1126/science.adg9210} {\bibfield  {journal} {\bibinfo  {journal} {Science}\ }\textbf {\bibinfo {volume} {381}},\ \bibinfo {pages} {221} (\bibinfo {year} {2023})}\BibitemShut {NoStop}%
\bibitem [{\citenamefont {Chi}\ \emph {et~al.}(2023)\citenamefont {Chi}, \citenamefont {Yu}, \citenamefont {Gong},\ and\ \citenamefont {Wang}}]{chi_high-dimensional_2023}%
  \BibitemOpen
  \bibfield  {author} {\bibinfo {author} {\bibfnamefont {Y.}~\bibnamefont {Chi}}, \bibinfo {author} {\bibfnamefont {Y.}~\bibnamefont {Yu}}, \bibinfo {author} {\bibfnamefont {Q.}~\bibnamefont {Gong}},\ and\ \bibinfo {author} {\bibfnamefont {J.}~\bibnamefont {Wang}},\ }\bibfield  {title} {\bibinfo {title} {High-dimensional quantum information processing on programmable integrated photonic chips},\ }\href {https://doi.org/10.1007/s11432-022-3602-0} {\bibfield  {journal} {\bibinfo  {journal} {Science China Information Sciences}\ }\textbf {\bibinfo {volume} {66}},\ \bibinfo {pages} {180501} (\bibinfo {year} {2023})}\BibitemShut {NoStop}%
\bibitem [{\citenamefont {Barreiro}\ \emph {et~al.}(2005)\citenamefont {Barreiro}, \citenamefont {Langford}, \citenamefont {Peters},\ and\ \citenamefont {Kwiat}}]{barreiro_generation_2005}%
  \BibitemOpen
  \bibfield  {author} {\bibinfo {author} {\bibfnamefont {J.~T.}\ \bibnamefont {Barreiro}}, \bibinfo {author} {\bibfnamefont {N.~K.}\ \bibnamefont {Langford}}, \bibinfo {author} {\bibfnamefont {N.~A.}\ \bibnamefont {Peters}},\ and\ \bibinfo {author} {\bibfnamefont {P.~G.}\ \bibnamefont {Kwiat}},\ }\bibfield  {title} {\bibinfo {title} {Generation of {Hyperentangled} {Photon} {Pairs}},\ }\href {https://doi.org/10.1103/PhysRevLett.95.260501} {\bibfield  {journal} {\bibinfo  {journal} {Phys. Rev. Lett.}\ }\textbf {\bibinfo {volume} {95}},\ \bibinfo {pages} {260501} (\bibinfo {year} {2005})}\BibitemShut {NoStop}%
\bibitem [{\citenamefont {Gao}\ \emph {et~al.}(2010)\citenamefont {Gao}, \citenamefont {Lu}, \citenamefont {Yao}, \citenamefont {Xu}, \citenamefont {Gühne}, \citenamefont {Goebel}, \citenamefont {Chen}, \citenamefont {Peng}, \citenamefont {Chen},\ and\ \citenamefont {Pan}}]{gao_experimental_2010}%
  \BibitemOpen
  \bibfield  {author} {\bibinfo {author} {\bibfnamefont {W.-B.}\ \bibnamefont {Gao}}, \bibinfo {author} {\bibfnamefont {C.-Y.}\ \bibnamefont {Lu}}, \bibinfo {author} {\bibfnamefont {X.-C.}\ \bibnamefont {Yao}}, \bibinfo {author} {\bibfnamefont {P.}~\bibnamefont {Xu}}, \bibinfo {author} {\bibfnamefont {O.}~\bibnamefont {Gühne}}, \bibinfo {author} {\bibfnamefont {A.}~\bibnamefont {Goebel}}, \bibinfo {author} {\bibfnamefont {Y.-A.}\ \bibnamefont {Chen}}, \bibinfo {author} {\bibfnamefont {C.-Z.}\ \bibnamefont {Peng}}, \bibinfo {author} {\bibfnamefont {Z.-B.}\ \bibnamefont {Chen}},\ and\ \bibinfo {author} {\bibfnamefont {J.-W.}\ \bibnamefont {Pan}},\ }\bibfield  {title} {\bibinfo {title} {Experimental demonstration of a hyper-entangled ten-qubit {Schrödinger} cat state},\ }\href {https://doi.org/10.1038/nphys1603} {\bibfield  {journal} {\bibinfo  {journal} {Nature Phys}\ }\textbf {\bibinfo {volume} {6}},\ \bibinfo {pages} {331} (\bibinfo {year} {2010})}\BibitemShut {NoStop}%
\bibitem [{\citenamefont {Vallone}\ \emph {et~al.}(2009)\citenamefont {Vallone}, \citenamefont {Ceccarelli}, \citenamefont {De~Martini},\ and\ \citenamefont {Mataloni}}]{vallone_hyperentanglement_2009}%
  \BibitemOpen
  \bibfield  {author} {\bibinfo {author} {\bibfnamefont {G.}~\bibnamefont {Vallone}}, \bibinfo {author} {\bibfnamefont {R.}~\bibnamefont {Ceccarelli}}, \bibinfo {author} {\bibfnamefont {F.}~\bibnamefont {De~Martini}},\ and\ \bibinfo {author} {\bibfnamefont {P.}~\bibnamefont {Mataloni}},\ }\bibfield  {title} {\bibinfo {title} {Hyperentanglement of two photons in three degrees of freedom},\ }\href {https://doi.org/10.1103/PhysRevA.79.030301} {\bibfield  {journal} {\bibinfo  {journal} {Phys. Rev. A}\ }\textbf {\bibinfo {volume} {79}},\ \bibinfo {pages} {030301} (\bibinfo {year} {2009})}\BibitemShut {NoStop}%
\bibitem [{\citenamefont {Ciampini}\ \emph {et~al.}(2016)\citenamefont {Ciampini}, \citenamefont {Orieux}, \citenamefont {Paesani}, \citenamefont {Sciarrino}, \citenamefont {Corrielli}, \citenamefont {Crespi}, \citenamefont {Ramponi}, \citenamefont {Osellame},\ and\ \citenamefont {Mataloni}}]{ciampini_path-polarization_2016}%
  \BibitemOpen
  \bibfield  {author} {\bibinfo {author} {\bibfnamefont {M.~A.}\ \bibnamefont {Ciampini}}, \bibinfo {author} {\bibfnamefont {A.}~\bibnamefont {Orieux}}, \bibinfo {author} {\bibfnamefont {S.}~\bibnamefont {Paesani}}, \bibinfo {author} {\bibfnamefont {F.}~\bibnamefont {Sciarrino}}, \bibinfo {author} {\bibfnamefont {G.}~\bibnamefont {Corrielli}}, \bibinfo {author} {\bibfnamefont {A.}~\bibnamefont {Crespi}}, \bibinfo {author} {\bibfnamefont {R.}~\bibnamefont {Ramponi}}, \bibinfo {author} {\bibfnamefont {R.}~\bibnamefont {Osellame}},\ and\ \bibinfo {author} {\bibfnamefont {P.}~\bibnamefont {Mataloni}},\ }\bibfield  {title} {\bibinfo {title} {Path-polarization hyperentangled and cluster states of photons on a chip},\ }\href {https://doi.org/10.1038/lsa.2016.64} {\bibfield  {journal} {\bibinfo  {journal} {Light Sci Appl}\ }\textbf {\bibinfo {volume} {5}},\ \bibinfo {pages} {e16064} (\bibinfo {year} {2016})}\BibitemShut {NoStop}%
\bibitem [{\citenamefont {Congia}\ \emph {et~al.}(2025)\citenamefont {Congia}, \citenamefont {Borghi}, \citenamefont {Brusaschi}, \citenamefont {Sabattoli}, \citenamefont {Dirani}, \citenamefont {Youssef}, \citenamefont {Pargon}, \citenamefont {Petit-Etienne}, \citenamefont {Sciancalepore}, \citenamefont {Liscidini}, \citenamefont {Rothman}, \citenamefont {Olivier}, \citenamefont {Galli},\ and\ \citenamefont {Bajoni}}]{congia_generation_2025}%
  \BibitemOpen
  \bibfield  {author} {\bibinfo {author} {\bibfnamefont {S.}~\bibnamefont {Congia}}, \bibinfo {author} {\bibfnamefont {M.}~\bibnamefont {Borghi}}, \bibinfo {author} {\bibfnamefont {E.}~\bibnamefont {Brusaschi}}, \bibinfo {author} {\bibfnamefont {F.~A.}\ \bibnamefont {Sabattoli}}, \bibinfo {author} {\bibfnamefont {H.~E.}\ \bibnamefont {Dirani}}, \bibinfo {author} {\bibfnamefont {L.}~\bibnamefont {Youssef}}, \bibinfo {author} {\bibfnamefont {E.}~\bibnamefont {Pargon}}, \bibinfo {author} {\bibfnamefont {C.}~\bibnamefont {Petit-Etienne}}, \bibinfo {author} {\bibfnamefont {C.}~\bibnamefont {Sciancalepore}}, \bibinfo {author} {\bibfnamefont {M.}~\bibnamefont {Liscidini}}, \bibinfo {author} {\bibfnamefont {J.}~\bibnamefont {Rothman}}, \bibinfo {author} {\bibfnamefont {S.}~\bibnamefont {Olivier}}, \bibinfo {author} {\bibfnamefont {M.}~\bibnamefont {Galli}},\ and\ \bibinfo {author} {\bibfnamefont {D.}~\bibnamefont {Bajoni}},\ }\bibfield  {title} {\bibinfo {title} {Generation of hyperentangled photon pairs in the time
  and frequency domain on a silicon photonic chip},\ }\href {https://doi.org/10.1364/OL.562079} {\bibfield  {journal} {\bibinfo  {journal} {Opt. Lett.}\ }\textbf {\bibinfo {volume} {50}},\ \bibinfo {pages} {5117} (\bibinfo {year} {2025})}\BibitemShut {NoStop}%
\bibitem [{\citenamefont {Wang}\ \emph {et~al.}(2023)\citenamefont {Wang}, \citenamefont {Sokolov},\ and\ \citenamefont {Agarwal}}]{wang_2023_quantum}%
  \BibitemOpen
  \bibfield  {author} {\bibinfo {author} {\bibfnamefont {J.}~\bibnamefont {Wang}}, \bibinfo {author} {\bibfnamefont {A.~V.}\ \bibnamefont {Sokolov}},\ and\ \bibinfo {author} {\bibfnamefont {G.~S.}\ \bibnamefont {Agarwal}},\ }\bibfield  {title} {\bibinfo {title} {Quantum entanglement between signal and frequency-up-converted idler photons},\ }\href {https://doi.org/10.1103/PhysRevA.108.063706} {\bibfield  {journal} {\bibinfo  {journal} {Phys. Rev. A}\ }\textbf {\bibinfo {volume} {108}},\ \bibinfo {pages} {063706} (\bibinfo {year} {2023})}\BibitemShut {NoStop}%
\bibitem [{\citenamefont {Takesue}\ and\ \citenamefont {Noguchi}(2009)}]{Takesue:09}%
  \BibitemOpen
  \bibfield  {author} {\bibinfo {author} {\bibfnamefont {H.}~\bibnamefont {Takesue}}\ and\ \bibinfo {author} {\bibfnamefont {Y.}~\bibnamefont {Noguchi}},\ }\bibfield  {title} {\bibinfo {title} {Implementation of quantum state tomography for time-bin entangled photon pairs},\ }\href {https://doi.org/10.1364/OE.17.010976} {\bibfield  {journal} {\bibinfo  {journal} {Opt. Express}\ }\textbf {\bibinfo {volume} {17}},\ \bibinfo {pages} {10976} (\bibinfo {year} {2009})}\BibitemShut {NoStop}%
\bibitem [{\citenamefont {Shahwar}\ \emph {et~al.}(2024)\citenamefont {Shahwar}, \citenamefont {Yoon}, \citenamefont {Akkanen}, \citenamefont {Li}, \citenamefont {Muntaha}, \citenamefont {Cherchi}, \citenamefont {Aalto},\ and\ \citenamefont {Sun}}]{shahwar_polarization_2024}%
  \BibitemOpen
  \bibfield  {author} {\bibinfo {author} {\bibfnamefont {D.}~\bibnamefont {Shahwar}}, \bibinfo {author} {\bibfnamefont {H.~H.}\ \bibnamefont {Yoon}}, \bibinfo {author} {\bibfnamefont {S.-T.}\ \bibnamefont {Akkanen}}, \bibinfo {author} {\bibfnamefont {D.}~\bibnamefont {Li}}, \bibinfo {author} {\bibfnamefont {S.~t.}\ \bibnamefont {Muntaha}}, \bibinfo {author} {\bibfnamefont {M.}~\bibnamefont {Cherchi}}, \bibinfo {author} {\bibfnamefont {T.}~\bibnamefont {Aalto}},\ and\ \bibinfo {author} {\bibfnamefont {Z.}~\bibnamefont {Sun}},\ }\bibfield  {title} {\bibinfo {title} {Polarization management in silicon photonics},\ }\href {https://doi.org/10.1038/s44310-024-00033-6} {\bibfield  {journal} {\bibinfo  {journal} {npj Nanophotonics}\ }\textbf {\bibinfo {volume} {1}},\ \bibinfo {pages} {35} (\bibinfo {year} {2024})}\BibitemShut {NoStop}%
\bibitem [{\citenamefont {Paesani}\ \emph {et~al.}(2020)\citenamefont {Paesani}, \citenamefont {Borghi}, \citenamefont {Signorini}, \citenamefont {Maïnos}, \citenamefont {Pavesi},\ and\ \citenamefont {Laing}}]{paesani_near-ideal_2020}%
  \BibitemOpen
  \bibfield  {author} {\bibinfo {author} {\bibfnamefont {S.}~\bibnamefont {Paesani}}, \bibinfo {author} {\bibfnamefont {M.}~\bibnamefont {Borghi}}, \bibinfo {author} {\bibfnamefont {S.}~\bibnamefont {Signorini}}, \bibinfo {author} {\bibfnamefont {A.}~\bibnamefont {Maïnos}}, \bibinfo {author} {\bibfnamefont {L.}~\bibnamefont {Pavesi}},\ and\ \bibinfo {author} {\bibfnamefont {A.}~\bibnamefont {Laing}},\ }\bibfield  {title} {\bibinfo {title} {Near-ideal spontaneous photon sources in silicon quantum photonics},\ }\bibfield  {journal} {\bibinfo  {journal} {Nature Communications}\ }\textbf {\bibinfo {volume} {11}},\ \href {https://doi.org/10.1038/s41467-020-16187-8} {10.1038/s41467-020-16187-8} (\bibinfo {year} {2020})\BibitemShut {NoStop}%
\bibitem [{sup()}]{supp}%
  \BibitemOpen
  \href@noop {} {}\bibinfo {note} {See Supplemental Material below for additional details on device characterisation and methods.}\BibitemShut {Stop}%
\bibitem [{\citenamefont {{Lumerical Inc.}}()}]{Lumerical}%
  \BibitemOpen
  \bibinfo {author} {\bibnamefont {{Lumerical Inc.}}}\BibitemShut {Stop}%
\bibitem [{\citenamefont {Silverstone}\ \emph {et~al.}(2014)\citenamefont {Silverstone}, \citenamefont {Bonneau}, \citenamefont {Ohira}, \citenamefont {Suzuki}, \citenamefont {Yoshida}, \citenamefont {Iizuka}, \citenamefont {Ezaki}, \citenamefont {Natarajan}, \citenamefont {Tanner}, \citenamefont {Hadfield}, \citenamefont {Zwiller}, \citenamefont {Marshall}, \citenamefont {Rarity}, \citenamefont {O'Brien},\ and\ \citenamefont {Thompson}}]{silverstone_2014}%
  \BibitemOpen
\bibfield  {author} {  }\bibfield  {author} {\bibinfo {author} {\bibfnamefont {J.~W.}\ \bibnamefont {Silverstone}}, \bibinfo {author} {\bibfnamefont {D.}~\bibnamefont {Bonneau}}, \bibinfo {author} {\bibfnamefont {K.}~\bibnamefont {Ohira}}, \bibinfo {author} {\bibfnamefont {N.}~\bibnamefont {Suzuki}}, \bibinfo {author} {\bibfnamefont {H.}~\bibnamefont {Yoshida}}, \bibinfo {author} {\bibfnamefont {N.}~\bibnamefont {Iizuka}}, \bibinfo {author} {\bibfnamefont {M.}~\bibnamefont {Ezaki}}, \bibinfo {author} {\bibfnamefont {C.~M.}\ \bibnamefont {Natarajan}}, \bibinfo {author} {\bibfnamefont {M.~G.}\ \bibnamefont {Tanner}}, \bibinfo {author} {\bibfnamefont {R.~H.}\ \bibnamefont {Hadfield}}, \bibinfo {author} {\bibfnamefont {V.}~\bibnamefont {Zwiller}}, \bibinfo {author} {\bibfnamefont {G.~D.}\ \bibnamefont {Marshall}}, \bibinfo {author} {\bibfnamefont {J.~G.}\ \bibnamefont {Rarity}}, \bibinfo {author} {\bibfnamefont {J.~L.}\ \bibnamefont {O'Brien}},\ and\ \bibinfo {author} {\bibfnamefont {M.~G.}\ \bibnamefont
  {Thompson}},\ }\bibfield  {title} {\bibinfo {title} {On-chip quantum interference between silicon photon-pair sources},\ }\href {https://doi.org/10.1038/nphoton.2013.339} {\bibfield  {journal} {\bibinfo  {journal} {Nature Photon}\ }\textbf {\bibinfo {volume} {8}},\ \bibinfo {pages} {104} (\bibinfo {year} {2014})}\BibitemShut {NoStop}%
\bibitem [{\citenamefont {Altepeter}\ \emph {et~al.}(2004)\citenamefont {Altepeter}, \citenamefont {James},\ and\ \citenamefont {Kwiat}}]{Altepeter2004}%
  \BibitemOpen
  \bibfield  {author} {\bibinfo {author} {\bibfnamefont {J.~B.}\ \bibnamefont {Altepeter}}, \bibinfo {author} {\bibfnamefont {D.~F.}\ \bibnamefont {James}},\ and\ \bibinfo {author} {\bibfnamefont {P.~G.}\ \bibnamefont {Kwiat}},\ }\bibinfo {title} {{Qubit} {Quantum} {State} {Tomography}},\ in\ \href@noop {} {\emph {\bibinfo {booktitle} {Quantum {State} {Estimation}}}},\ \bibinfo {editor} {edited by\ \bibinfo {editor} {\bibfnamefont {M.}~\bibnamefont {Paris}}\ and\ \bibinfo {editor} {\bibfnamefont {J.}~\bibnamefont {Řeháček}}}\ (\bibinfo  {publisher} {Springer Berlin},\ \bibinfo {address} {Heidelberg},\ \bibinfo {year} {2004})\ Chap.~\bibinfo {chapter} {4}, pp.\ \bibinfo {pages} {113--145},\ \bibinfo {edition} {1st}\ ed.\BibitemShut {Stop}%
\bibitem [{\citenamefont {Hradil}\ \emph {et~al.}(2004{\natexlab{a}})\citenamefont {Hradil}, \citenamefont {Řeháček}, \citenamefont {Fiurášek},\ and\ \citenamefont {Ježek}}]{Hradil2004}%
  \BibitemOpen
  \bibfield  {author} {\bibinfo {author} {\bibfnamefont {Z.}~\bibnamefont {Hradil}}, \bibinfo {author} {\bibfnamefont {J.}~\bibnamefont {Řeháček}}, \bibinfo {author} {\bibfnamefont {J.}~\bibnamefont {Fiurášek}},\ and\ \bibinfo {author} {\bibfnamefont {M.}~\bibnamefont {Ježek}},\ }\bibinfo {title} {{Maximum}-{Likelihood} {Methods} in {Quantum} {Mechanics}},\ in\ \href@noop {} {\emph {\bibinfo {booktitle} {Quantum {State} {Estimation}}}},\ \bibinfo {editor} {edited by\ \bibinfo {editor} {\bibfnamefont {M.}~\bibnamefont {Paris}}\ and\ \bibinfo {editor} {\bibfnamefont {J.}~\bibnamefont {Řeháček}}}\ (\bibinfo  {publisher} {Springer Berlin},\ \bibinfo {address} {Heidelberg},\ \bibinfo {year} {2004})\ Chap.~\bibinfo {chapter} {3}, pp.\ \bibinfo {pages} {59--112},\ \bibinfo {edition} {1st}\ ed.\BibitemShut {Stop}%
\bibitem [{\citenamefont {Nielsen}\ and\ \citenamefont {Chuang}(2010)}]{nielsen_quantum_2010}%
  \BibitemOpen
  \bibfield  {author} {\bibinfo {author} {\bibfnamefont {M.~A.}\ \bibnamefont {Nielsen}}\ and\ \bibinfo {author} {\bibfnamefont {I.~L.}\ \bibnamefont {Chuang}},\ }\href@noop {} {\emph {\bibinfo {title} {Quantum Computation and Quantum Information}}},\ \bibinfo {edition} {10th}\ ed.\ (\bibinfo  {publisher} {Cambridge University Press},\ \bibinfo {address} {Cambridge},\ \bibinfo {year} {2010})\BibitemShut {NoStop}%
\bibitem [{\citenamefont {Schlosshauer}(2019)}]{SCHLOSSHAUER20191}%
  \BibitemOpen
  \bibfield  {author} {\bibinfo {author} {\bibfnamefont {M.}~\bibnamefont {Schlosshauer}},\ }\bibfield  {title} {\bibinfo {title} {Quantum decoherence},\ }\href {https://doi.org/https://doi.org/10.1016/j.physrep.2019.10.001} {\bibfield  {journal} {\bibinfo  {journal} {Physics Reports}\ }\textbf {\bibinfo {volume} {831}},\ \bibinfo {pages} {1} (\bibinfo {year} {2019})}\BibitemShut {NoStop}%
\bibitem [{\citenamefont {Yan}\ \emph {et~al.}(2023)\citenamefont {Yan}, \citenamefont {Zhou}, \citenamefont {Zhong},\ and\ \citenamefont {Sheng}}]{yan_entanglementpurification_2023}%
  \BibitemOpen
  \bibfield  {author} {\bibinfo {author} {\bibfnamefont {P.-S.}\ \bibnamefont {Yan}}, \bibinfo {author} {\bibfnamefont {L.}~\bibnamefont {Zhou}}, \bibinfo {author} {\bibfnamefont {W.}~\bibnamefont {Zhong}},\ and\ \bibinfo {author} {\bibfnamefont {Y.-B.}\ \bibnamefont {Sheng}},\ }\bibfield  {title} {\bibinfo {title} {Advances in quantum entanglement purification},\ }\bibfield  {journal} {\bibinfo  {journal} {Science China Physics, Mechanics \& Astronomy}\ }\textbf {\bibinfo {volume} {66}},\ \href {https://doi.org/10.1007/s11433-022-2065-x} {10.1007/s11433-022-2065-x} (\bibinfo {year} {2023})\BibitemShut {NoStop}%
\bibitem [{\citenamefont {Faurby}\ \emph {et~al.}(2024)\citenamefont {Faurby}, \citenamefont {Carosini}, \citenamefont {Cao}, \citenamefont {Sund}, \citenamefont {Hansen}, \citenamefont {Giorgino}, \citenamefont {Villadsen}, \citenamefont {van~den Hoven}, \citenamefont {Lodahl}, \citenamefont {Paesani}, \citenamefont {Loredo},\ and\ \citenamefont {Walther}}]{faurby_purifying_2024}%
  \BibitemOpen
  \bibfield  {author} {\bibinfo {author} {\bibfnamefont {C.~F.}\ \bibnamefont {Faurby}}, \bibinfo {author} {\bibfnamefont {L.}~\bibnamefont {Carosini}}, \bibinfo {author} {\bibfnamefont {H.}~\bibnamefont {Cao}}, \bibinfo {author} {\bibfnamefont {P.~I.}\ \bibnamefont {Sund}}, \bibinfo {author} {\bibfnamefont {L.~M.}\ \bibnamefont {Hansen}}, \bibinfo {author} {\bibfnamefont {F.}~\bibnamefont {Giorgino}}, \bibinfo {author} {\bibfnamefont {A.~B.}\ \bibnamefont {Villadsen}}, \bibinfo {author} {\bibfnamefont {S.~N.}\ \bibnamefont {van~den Hoven}}, \bibinfo {author} {\bibfnamefont {P.}~\bibnamefont {Lodahl}}, \bibinfo {author} {\bibfnamefont {S.}~\bibnamefont {Paesani}}, \bibinfo {author} {\bibfnamefont {J.~C.}\ \bibnamefont {Loredo}},\ and\ \bibinfo {author} {\bibfnamefont {P.}~\bibnamefont {Walther}},\ }\bibfield  {title} {\bibinfo {title} {Purifying {Photon} {Indistinguishability} through {Quantum} {Interference}},\ }\href {https://doi.org/10.1103/PhysRevLett.133.033604} {\bibfield  {journal} {\bibinfo  {journal}
  {Phys. Rev. Lett.}\ }\textbf {\bibinfo {volume} {133}},\ \bibinfo {pages} {033604} (\bibinfo {year} {2024})}\BibitemShut {NoStop}%
\bibitem [{\citenamefont {Hoch}\ \emph {et~al.}(2025)\citenamefont {Hoch}, \citenamefont {Camillini}, \citenamefont {Rodari}, \citenamefont {Caruccio}, \citenamefont {Carvacho}, \citenamefont {Giordani}, \citenamefont {Albiero}, \citenamefont {Giano}, \citenamefont {Corrielli}, \citenamefont {Ceccarelli}, \citenamefont {Osellame}, \citenamefont {Robbio}, \citenamefont {Novo}, \citenamefont {Spagnolo}, \citenamefont {Galvão},\ and\ \citenamefont {Sciarrino}}]{hoch_optimal_2025}%
  \BibitemOpen
  \bibfield  {author} {\bibinfo {author} {\bibfnamefont {F.}~\bibnamefont {Hoch}}, \bibinfo {author} {\bibfnamefont {A.}~\bibnamefont {Camillini}}, \bibinfo {author} {\bibfnamefont {G.}~\bibnamefont {Rodari}}, \bibinfo {author} {\bibfnamefont {E.}~\bibnamefont {Caruccio}}, \bibinfo {author} {\bibfnamefont {G.}~\bibnamefont {Carvacho}}, \bibinfo {author} {\bibfnamefont {T.}~\bibnamefont {Giordani}}, \bibinfo {author} {\bibfnamefont {R.}~\bibnamefont {Albiero}}, \bibinfo {author} {\bibfnamefont {N.~D.}\ \bibnamefont {Giano}}, \bibinfo {author} {\bibfnamefont {G.}~\bibnamefont {Corrielli}}, \bibinfo {author} {\bibfnamefont {F.}~\bibnamefont {Ceccarelli}}, \bibinfo {author} {\bibfnamefont {R.}~\bibnamefont {Osellame}}, \bibinfo {author} {\bibfnamefont {M.}~\bibnamefont {Robbio}}, \bibinfo {author} {\bibfnamefont {L.}~\bibnamefont {Novo}}, \bibinfo {author} {\bibfnamefont {N.}~\bibnamefont {Spagnolo}}, \bibinfo {author} {\bibfnamefont {E.~F.}\ \bibnamefont {Galvão}},\ and\ \bibinfo {author} {\bibfnamefont
  {F.}~\bibnamefont {Sciarrino}},\ }\href {https://doi.org/10.48550/arXiv.2509.02296} {\bibinfo {title} {Optimal distillation of photonic indistinguishability}} (\bibinfo {year} {2025}),\ \bibinfo {note} {arXiv:2509.02296 [quant-ph]}\BibitemShut {NoStop}%
\bibitem [{\citenamefont {Bennett}\ \emph {et~al.}(1996)\citenamefont {Bennett}, \citenamefont {Brassard}, \citenamefont {Popescu}, \citenamefont {Schumacher}, \citenamefont {Smolin},\ and\ \citenamefont {Wootters}}]{bennett_purification_1996}%
  \BibitemOpen
  \bibfield  {author} {\bibinfo {author} {\bibfnamefont {C.~H.}\ \bibnamefont {Bennett}}, \bibinfo {author} {\bibfnamefont {G.}~\bibnamefont {Brassard}}, \bibinfo {author} {\bibfnamefont {S.}~\bibnamefont {Popescu}}, \bibinfo {author} {\bibfnamefont {B.}~\bibnamefont {Schumacher}}, \bibinfo {author} {\bibfnamefont {J.~A.}\ \bibnamefont {Smolin}},\ and\ \bibinfo {author} {\bibfnamefont {W.~K.}\ \bibnamefont {Wootters}},\ }\bibfield  {title} {\bibinfo {title} {Purification of {Noisy} {Entanglement} and {Faithful} {Teleportation} via {Noisy} {Channels}},\ }\href {https://doi.org/10.1103/PhysRevLett.76.722} {\bibfield  {journal} {\bibinfo  {journal} {Phys. Rev. Lett.}\ }\textbf {\bibinfo {volume} {76}},\ \bibinfo {pages} {722} (\bibinfo {year} {1996})}\BibitemShut {NoStop}%
\bibitem [{\citenamefont {Slussarenko}\ and\ \citenamefont {Pryde}(2019)}]{Slussarenko_Photoic_2019}%
  \BibitemOpen
  \bibfield  {author} {\bibinfo {author} {\bibfnamefont {S.}~\bibnamefont {Slussarenko}}\ and\ \bibinfo {author} {\bibfnamefont {G.~J.}\ \bibnamefont {Pryde}},\ }\bibfield  {title} {\bibinfo {title} {Photonic quantum information processing: A concise review},\ }\href {https://doi.org/10.1063/1.5115814} {\bibfield  {journal} {\bibinfo  {journal} {Applied Physics Reviews}\ }\textbf {\bibinfo {volume} {6}},\ \bibinfo {pages} {041303} (\bibinfo {year} {2019})}\BibitemShut {NoStop}%
\bibitem [{\citenamefont {Pittman}\ \emph {et~al.}(2003)\citenamefont {Pittman}, \citenamefont {Fitch}, \citenamefont {Jacobs},\ and\ \citenamefont {Franson}}]{pittman_experimental_2003}%
  \BibitemOpen
  \bibfield  {author} {\bibinfo {author} {\bibfnamefont {T.~B.}\ \bibnamefont {Pittman}}, \bibinfo {author} {\bibfnamefont {M.~J.}\ \bibnamefont {Fitch}}, \bibinfo {author} {\bibfnamefont {B.~C.}\ \bibnamefont {Jacobs}},\ and\ \bibinfo {author} {\bibfnamefont {J.~D.}\ \bibnamefont {Franson}},\ }\bibfield  {title} {\bibinfo {title} {Experimental controlled-not logic gate for single photons in the coincidence basis},\ }\href {https://doi.org/10.1103/PhysRevA.68.032316} {\bibfield  {journal} {\bibinfo  {journal} {Phys. Rev. A}\ }\textbf {\bibinfo {volume} {68}},\ \bibinfo {pages} {032316} (\bibinfo {year} {2003})}\BibitemShut {NoStop}%
\bibitem [{\citenamefont {Simon}\ and\ \citenamefont {Pan}(2002)}]{simon_polarization_2002}%
  \BibitemOpen
  \bibfield  {author} {\bibinfo {author} {\bibfnamefont {C.}~\bibnamefont {Simon}}\ and\ \bibinfo {author} {\bibfnamefont {J.-W.}\ \bibnamefont {Pan}},\ }\bibfield  {title} {\bibinfo {title} {Polarization entanglement purification using spatial entanglement},\ }\href {https://doi.org/10.1103/PhysRevLett.89.257901} {\bibfield  {journal} {\bibinfo  {journal} {Phys. Rev. Lett.}\ }\textbf {\bibinfo {volume} {89}},\ \bibinfo {pages} {257901} (\bibinfo {year} {2002})}\BibitemShut {NoStop}%
\bibitem [{\citenamefont {Ecker}\ \emph {et~al.}(2021)\citenamefont {Ecker}, \citenamefont {Sohr}, \citenamefont {Bulla}, \citenamefont {Huber}, \citenamefont {Bohmann},\ and\ \citenamefont {Ursin}}]{ecker_experimental_2021}%
  \BibitemOpen
  \bibfield  {author} {\bibinfo {author} {\bibfnamefont {S.}~\bibnamefont {Ecker}}, \bibinfo {author} {\bibfnamefont {P.}~\bibnamefont {Sohr}}, \bibinfo {author} {\bibfnamefont {L.}~\bibnamefont {Bulla}}, \bibinfo {author} {\bibfnamefont {M.}~\bibnamefont {Huber}}, \bibinfo {author} {\bibfnamefont {M.}~\bibnamefont {Bohmann}},\ and\ \bibinfo {author} {\bibfnamefont {R.}~\bibnamefont {Ursin}},\ }\bibfield  {title} {\bibinfo {title} {Experimental {Single}-{Copy} {Entanglement} {Distillation}},\ }\href {https://doi.org/10.1103/PhysRevLett.127.040506} {\bibfield  {journal} {\bibinfo  {journal} {Phys. Rev. Lett.}\ }\textbf {\bibinfo {volume} {127}},\ \bibinfo {pages} {040506} (\bibinfo {year} {2021})}\BibitemShut {NoStop}%
\bibitem [{\citenamefont {Kok}\ and\ \citenamefont {Lovett}(2010)}]{kok_qip_2010}%
  \BibitemOpen
  \bibfield  {author} {\bibinfo {author} {\bibfnamefont {P.}~\bibnamefont {Kok}}\ and\ \bibinfo {author} {\bibfnamefont {B.~W.}\ \bibnamefont {Lovett}},\ }\href@noop {} {\emph {\bibinfo {title} {Introduction to Optical Quantum Information Processing}}}\ (\bibinfo  {publisher} {Cambridge University Press},\ \bibinfo {address} {Cambridge},\ \bibinfo {year} {2010})\BibitemShut {NoStop}%
\bibitem [{\citenamefont {Paesani}(2019)}]{paesani_2019}%
  \BibitemOpen
  \bibfield  {author} {\bibinfo {author} {\bibfnamefont {S.}~\bibnamefont {Paesani}},\ }\emph {\bibinfo {title} {Large-Scale Integrated Quantum Photonics: Development and Applications}},\ \href@noop {} {\bibinfo {type} {{Ph.D.} thesis}},\ \bibinfo  {school} {University of Bristol} (\bibinfo {year} {2019})\BibitemShut {NoStop}%
\bibitem [{\citenamefont {Sych}\ \emph {et~al.}(2012)\citenamefont {Sych}, \citenamefont {{\ifmmode\check{R}\else\v{R}\fi}eh{\ifmmode\acute{a}\else\'{a}\fi}{\ifmmode\check{c}\else\v{c}\fi}ek}, \citenamefont {Hradil}, \citenamefont {Leuchs},\ and\ \citenamefont {S{\ifmmode\acute{a}\else\'{a}\fi}nchez-Soto}}]{Sych2012Nov}%
  \BibitemOpen
  \bibfield  {author} {\bibinfo {author} {\bibfnamefont {D.}~\bibnamefont {Sych}}, \bibinfo {author} {\bibfnamefont {J.}~\bibnamefont {{\ifmmode\check{R}\else\v{R}\fi}eh{\ifmmode\acute{a}\else\'{a}\fi}{\ifmmode\check{c}\else\v{c}\fi}ek}}, \bibinfo {author} {\bibfnamefont {Z.}~\bibnamefont {Hradil}}, \bibinfo {author} {\bibfnamefont {G.}~\bibnamefont {Leuchs}},\ and\ \bibinfo {author} {\bibfnamefont {L.~L.}\ \bibnamefont {S{\ifmmode\acute{a}\else\'{a}\fi}nchez-Soto}},\ }\bibfield  {title} {\bibinfo {title} {{Informational completeness of continuous-variable measurements}},\ }\href {https://doi.org/10.1103/PhysRevA.86.052123} {\bibfield  {journal} {\bibinfo  {journal} {Phys. Rev. A}\ }\textbf {\bibinfo {volume} {86}},\ \bibinfo {pages} {052123} (\bibinfo {year} {2012})}\BibitemShut {NoStop}%
\bibitem [{\citenamefont {{\ifmmode\check{R}\else\v{R}\fi}eh{\ifmmode\acute{a}\else\'{a}\fi}{\ifmmode\check{c}\else\v{c}\fi}ek}\ \emph {et~al.}(2009)\citenamefont {{\ifmmode\check{R}\else\v{R}\fi}eh{\ifmmode\acute{a}\else\'{a}\fi}{\ifmmode\check{c}\else\v{c}\fi}ek}, \citenamefont {Hradil}, \citenamefont {Bouchal}, \citenamefont {{\ifmmode\check{C}\else\v{C}\fi}elechovsk{\ifmmode\acute{y}\else\'{y}\fi}}, \citenamefont {Rigas},\ and\ \citenamefont {S{\ifmmode\acute{a}\else\'{a}\fi}nchez-Soto}}]{Rehacek2009Dec}%
  \BibitemOpen
  \bibfield  {author} {\bibinfo {author} {\bibfnamefont {J.}~\bibnamefont {{\ifmmode\check{R}\else\v{R}\fi}eh{\ifmmode\acute{a}\else\'{a}\fi}{\ifmmode\check{c}\else\v{c}\fi}ek}}, \bibinfo {author} {\bibfnamefont {Z.}~\bibnamefont {Hradil}}, \bibinfo {author} {\bibfnamefont {Z.}~\bibnamefont {Bouchal}}, \bibinfo {author} {\bibfnamefont {R.}~\bibnamefont {{\ifmmode\check{C}\else\v{C}\fi}elechovsk{\ifmmode\acute{y}\else\'{y}\fi}}}, \bibinfo {author} {\bibfnamefont {I.}~\bibnamefont {Rigas}},\ and\ \bibinfo {author} {\bibfnamefont {L.~L.}\ \bibnamefont {S{\ifmmode\acute{a}\else\'{a}\fi}nchez-Soto}},\ }\bibfield  {title} {\bibinfo {title} {{Full Tomography from Compatible Measurements}},\ }\href {https://doi.org/10.1103/PhysRevLett.103.250402} {\bibfield  {journal} {\bibinfo  {journal} {Phys. Rev. Lett.}\ }\textbf {\bibinfo {volume} {103}},\ \bibinfo {pages} {250402} (\bibinfo {year} {2009})}\BibitemShut {NoStop}%
\bibitem [{\citenamefont {Hradil}\ \emph {et~al.}(2004{\natexlab{b}})\citenamefont {Hradil}, \citenamefont {{\ifmmode\check{R}\else\v{R}\fi}eh{\ifmmode\acute{a}\else\'{a}\fi}{\ifmmode\check{c}\else\v{c}\fi}ek}, \citenamefont {Fiur{\ifmmode\acute{a}\else\'{a}\fi}{\ifmmode\check{s}\else\v{s}\fi}ek},\ and\ \citenamefont {Je{\ifmmode\check{z}\else\v{z}\fi}ek}}]{Hradil2004Aug}%
  \BibitemOpen
  \bibfield  {author} {\bibinfo {author} {\bibfnamefont {Z.}~\bibnamefont {Hradil}}, \bibinfo {author} {\bibfnamefont {J.}~\bibnamefont {{\ifmmode\check{R}\else\v{R}\fi}eh{\ifmmode\acute{a}\else\'{a}\fi}{\ifmmode\check{c}\else\v{c}\fi}ek}}, \bibinfo {author} {\bibfnamefont {J.}~\bibnamefont {Fiur{\ifmmode\acute{a}\else\'{a}\fi}{\ifmmode\check{s}\else\v{s}\fi}ek}},\ and\ \bibinfo {author} {\bibfnamefont {M.}~\bibnamefont {Je{\ifmmode\check{z}\else\v{z}\fi}ek}},\ }\bibfield  {title} {\bibinfo {title} {{3 Maximum-Likelihood Methodsin Quantum Mechanics}},\ }in\ \href {https://doi.org/10.1007/978-3-540-44481-7_3} {\emph {\bibinfo {booktitle} {{Quantum State Estimation}}}}\ (\bibinfo  {publisher} {Springer},\ \bibinfo {address} {Berlin, Germany},\ \bibinfo {year} {2004})\ pp.\ \bibinfo {pages} {59--112}\BibitemShut {NoStop}%
\end{thebibliography}%

\pagebreak
\clearpage

\title{Supplemental Material - \TitleName}
\maketitle
\onecolumngrid

\setcounter{equation}{0}
\setcounter{figure}{0}
\setcounter{table}{0}
\setcounter{section}{0}
\setcounter{page}{1}
\makeatletter
\renewcommand{\theequation}{S\arabic{equation}}
\renewcommand{\thefigure}{S\arabic{figure}}

\section{Experimental Setup}
\begin{figure}
\centering
\includegraphics {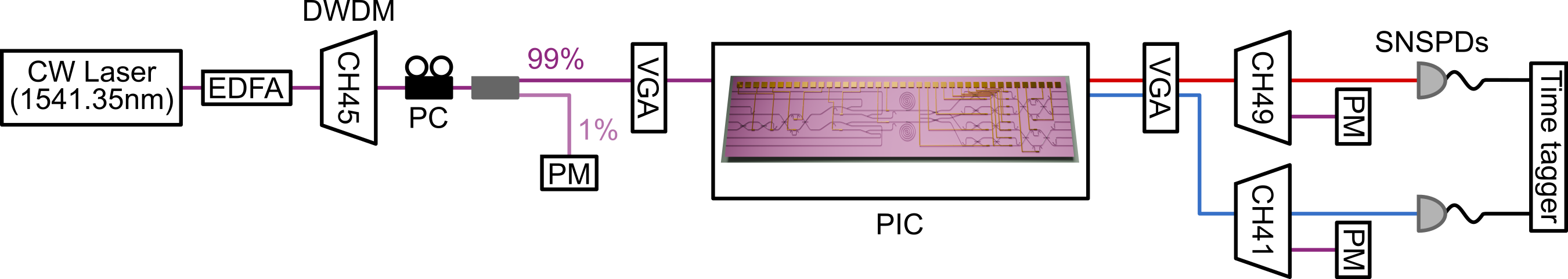}
\caption{\label{fig:exp_supp} Schematic showing the full experimental setup. EDFA: Erbium-Doped Fiber Amplifier, DWDM CHX: Dense Wavelength-Division Multiplexing ITU grid Channel X, PC: Polarization Controller, PM: Power Meter, VGA: V-Groove Array, PIC: Photonic Integrated Circuit, SNSPDs: Superconducting Nanowire Single Photon Detectors. }
\end{figure}

A schematic of the experimental setup used is given in Figure \ref{fig:exp_supp}. A fiber coupled, TUNICS T100 tunable continuous wave laser at 1541.35 nm (DWDM ITU grid Channel 45) is amplified using a PriTel EDFA. This pump light is then filtered using a single DWDM channel fiber filter (Channel 45) to suppress laser sidebands. A FibrePro in-line polarization controller is used to maximise coupling to the on-chip grating couplers. Then a 99:1 fiber tap is used to monitor the power of the pump light going to the chip by using a Thorlabs PM100D power meter on the 1\% output. The pump light is coupled on and off the chip via an Oz Optics VGA and grating couplers. Voltages for phase shifters on the chip are controlled using Qontrol Q8 boards and electronics. The chip is used to generate and measure quantum states as described in the main text, before coupling the photons off-chip. The signal and idler photons, at wavelengths of 1538.19 nm and 1544.53 nm, are then isolated and filtered from the pump light using a second set of DWDM filters (Channel 41 and 49). The pump light is incident on additional power meters to allow for monitoring. The signal and idler photons are then incident on Photonspot SNSPDs, and coincidence measurements are time tagged using a Swabian Instruments Time Tagger Ultra.

\section{Chip Component Characterization}
\begin{figure}
\centering
\includegraphics[scale = 0.5]{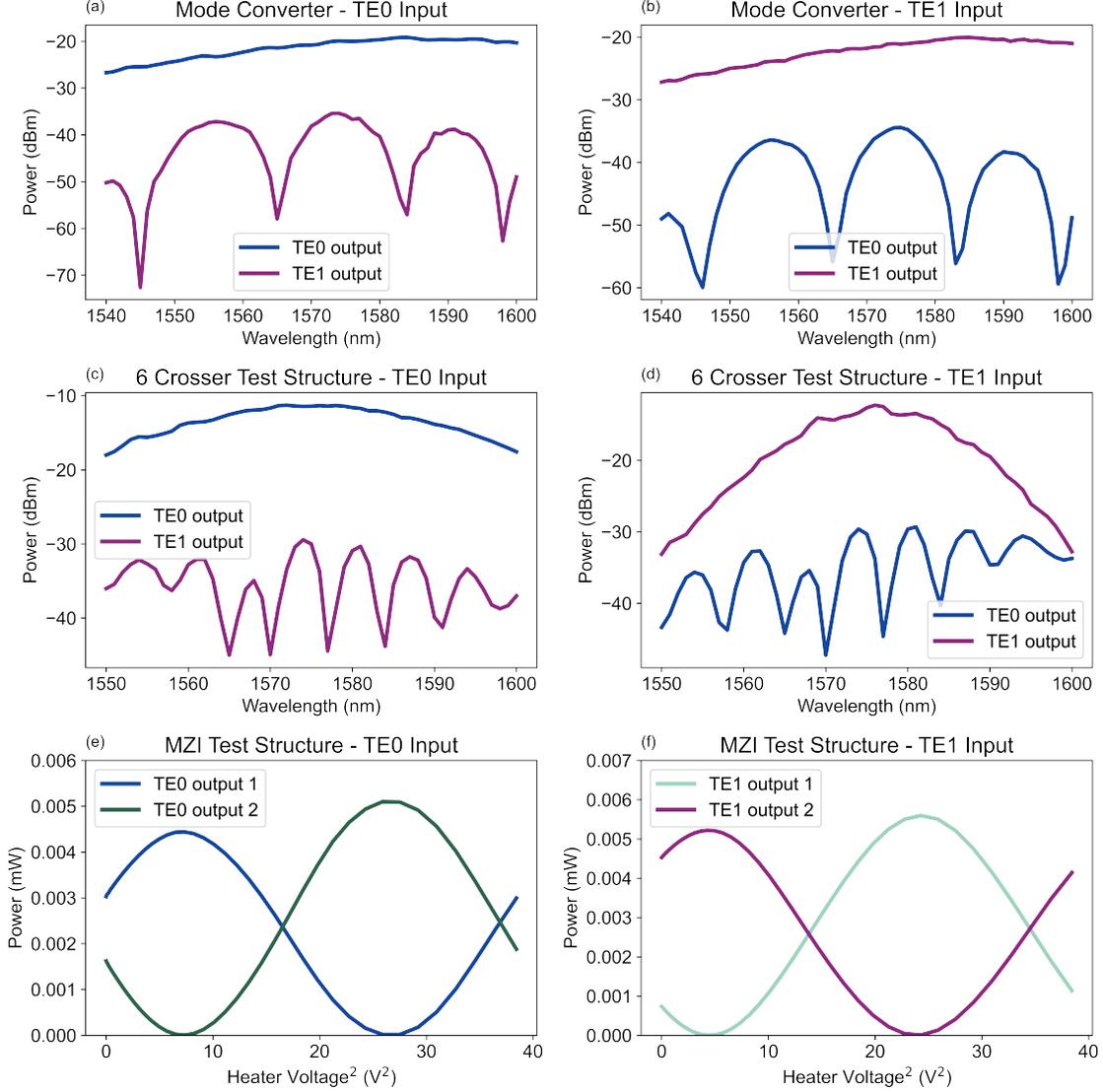}
\caption{\label{fig:characterisation}(a) - (f) Characterization data for the TE mode converter, multimode crossers and MZIs on the device. (a,b) TE0 and TE1 outputs for a mode converter test structure for a range of wavelengths with a TE0 or TE1 input. (c,d) Transmission through a test structure with six crossers on a single waveguide for TE0 and TE1 inputs, respectively. (e,f) Output power for a MZI with varying voltage applied to the heater, for a TE0 and TE1 input respectively.}
\end{figure}

Initial device characterization included characterizing the different intermodal structures on the device, such as mode converters, crossers, multimode interference (MMIs) couplers and Mach–Zehnder interferometers (MZIs). Results from this characterization are shown in Figure \ref{fig:characterisation}. 
Figure \ref{fig:characterisation}(a,b) show results for a mode converter structure where light is coupled from single mode waveguides into a multimode waveguide then back to single mode waveguide so that the individual modes can be coupled out and measured. We see minimal spurious mode conversion for these structures. For example, considering Channel 41, we see an extinction ratio for TE0 light of 40.1 dB, and an extinction ratio for TE1 light of 29.7 dB.  Figure \ref{fig:characterisation}(c,d) show transmission through a series of six multimodal waveguide crosser structures. Again, we see minimal spurious mode conversion during crosser propagation, however our extinction ratios here are smaller than for the mode converters. For Figure \ref{fig:characterisation}(a,b,c,d) we see a higher loss for TE1 light as opposed to TE0 light. This is likely due to less of the TE1 mode being confined in the waveguide structure, as for TE0 light our multimode waveguide has an effective index of 2.69, whereas for TE1 light is has an effective index of 2.18. Figure \ref{fig:characterisation}(e,f) show measured sinusoidal fringes in optical power, as expected when sweeping over a range of heater voltages for a phase shifter within an MZI structure. The visibilities for these fringes are all $> 99\%$.

\begin{figure}
\centering
\includegraphics[scale = 0.45]{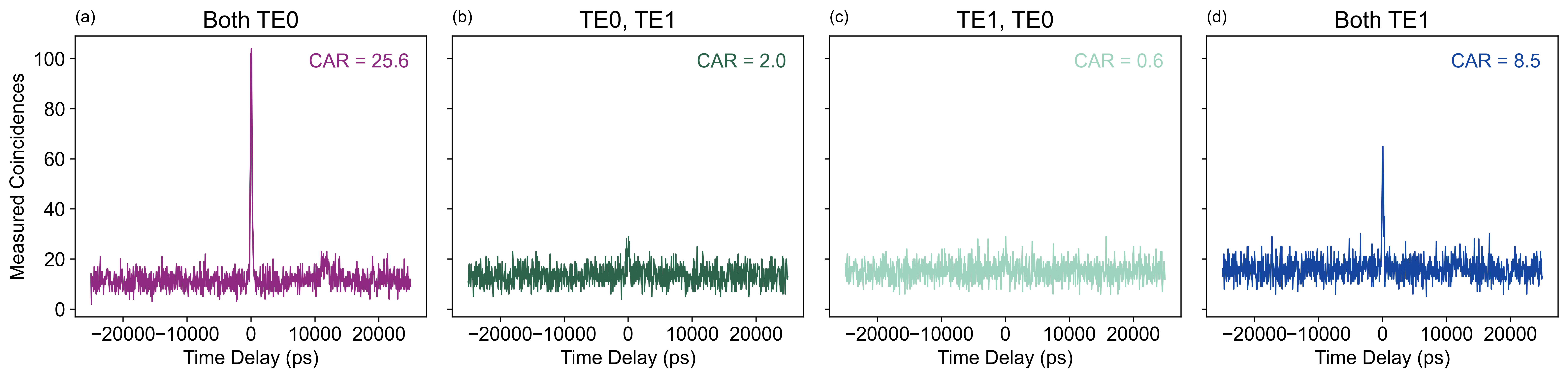}
\caption{\label{fig:source_characterisation} Coincidence peaks measured for intermodal SFWM, looking at correlations between (a) TE0 signal and idler outputs, (b) TE0 signal and TE1 idler, (c) TE1 signal and TE0 idler, and (d) TE1 signal and idler outputs. The coincidence-to-accidental ratio (CAR) is given on each plot.}
\end{figure}

For the spiral waveguide used in spontaneous four wave mixing (SFWM), the sources are designed to support pair generation in either the TE0 or TE1 modes, with minimal generation of pairs mixed in TE0 and TE1. To verify this behaviour, we pump both sources with an equal superposition of TE0 and TE1 light, then set all other interferometry on the chip to identity. We look at coincidence measurements for different combinations of TE0 and TE1 outputs on the chip, keeping all powers, integration times and phases set on the chip constant. The coincidence peaks in Figure \ref{fig:source_characterisation}(a,d) highlight the pair generation in the same mode, albeit at a lower rate for TE1 due to the higher losses for TE1 light throughout the chip. The lack of significant coincidence peaks and lower CAR values in Figure \ref{fig:source_characterisation}(b,c) shows there is minimal intermodal generation. The minimal peak that is present can, in part, be attributed to imperfect mode conversion in the multimodal components, as opposed to necessarily being from intermodal SFWM.

\section{Time-Reversed Hong-Ou-Mandel Effect (RHOM)}
\begin{figure}
\centering
\includegraphics[scale = 0.5] {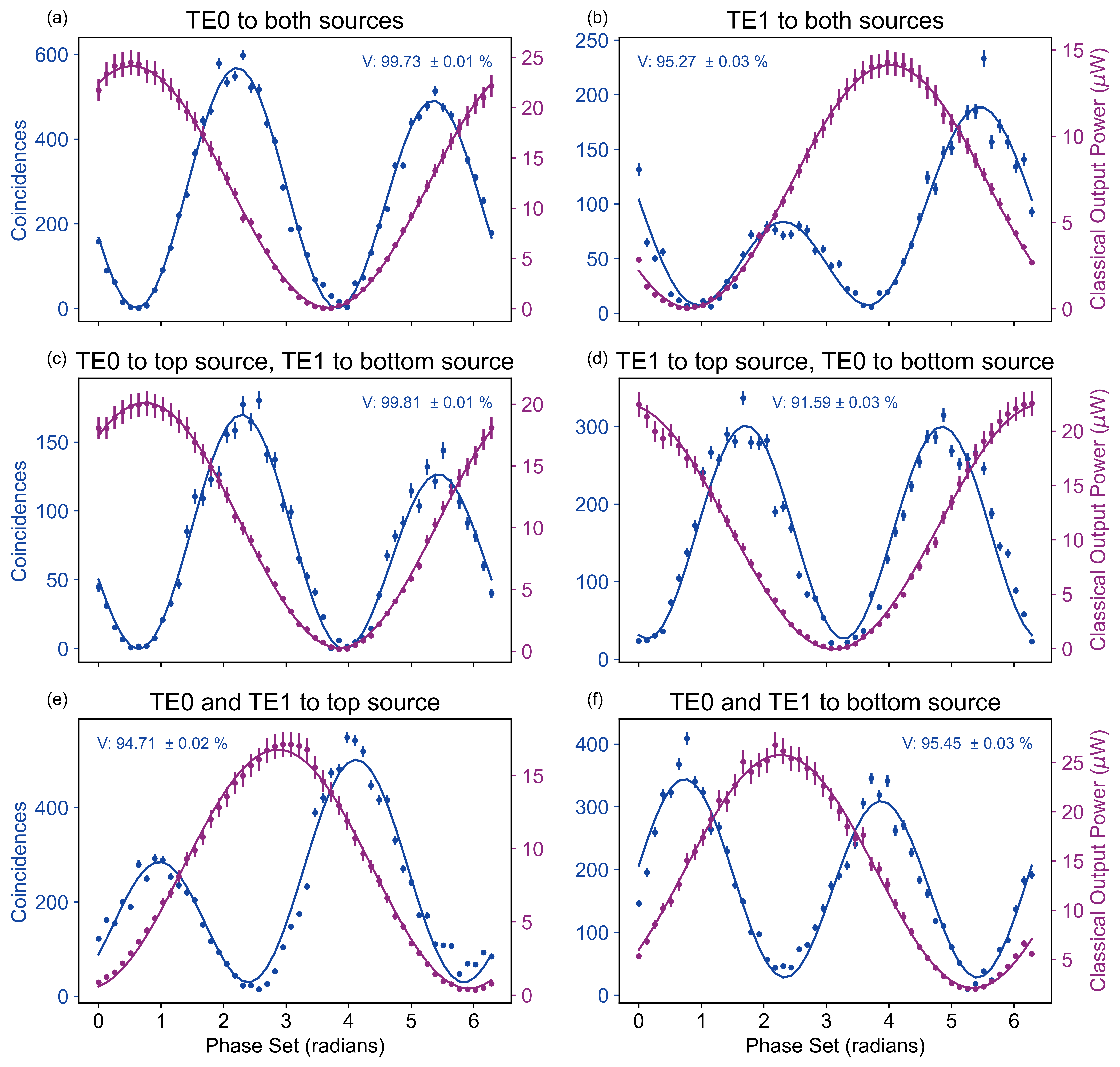}
\caption{\label{fig:revhom}Plots (a) - (f) show the fringes measured using a RHOM setup for different combinations of the two on-chip spiral sources, and for sending combinations of TE0 and TE1 light to each. The purple line measures the light from the pump laser travelling through the chip, and errors arise from the measurement uncertainty of the power sensor used. The blue line measures the coincidence counts measured between the two arms of the interferometer where we interfere our photons. Coincidence count measurements have background noise removed and the error bars are calculated using Poissonian statistics. The visibility, V, of the coincidence fringe is labelled on each plot, with errors from bootstrapping.}
\end{figure}

 To quantify the indistinguishability of photons from each source, we measure a RHOM fringe. We begin with the superposition $\frac{1}{\sqrt{2}} \ket{20} + \ket{02}$, in the Fock basis, and then apply a phase shift to the bottom mode. The phase in the superposition state gains an additional factor of two, as the phase shifter acts on the two photons. The two modes are interfered on a 50:50 beamsplitter and coincidence measurements are taken between the two output modes. The output state, up to a global phase and post selecting on single pair emission, can be written as:

\begin{equation}{\label{rev_hom_output_state}}
\ket{\psi_{out}} = \frac{1}{\sqrt{2}} \left( \text{cos}\left(\phi\right) \ket{11} - \text{sin}\left(\phi \right) \left(\ket{20} - \ket{02} \right) \right),
\end{equation}
where $\phi$ is the phase applied on the phase shifter. The probability of detecting a coincidence between the two output modes then varies as $p_{coinc} = \text{cos}^2\left(\phi \right)$. For classical light, the power seen on one of the output modes varies proportionally as $\text{cos}^2\left(\frac{\phi}{2}\right)$ \cite{silverstone_2014, paesani_2019}.

These interference effects are observed by sweeping over a range of phases, and measuring both the coincidence counts from the generated photons for the quantum case and power of the classical light through the chip for the classical fringe. The results of this are shown in Figure \ref{fig:revhom}, and show the frequency of the fringes from the coincidences to be double that of the classical light, as expected.

In Figure \ref{fig:revhom}, there is some asymmetry between the double-peaked fringes we see in the coincidence counts. We suggest that this is due to potential SFWM noise elsewhere in the chip, similar to the effect seen in \cite{silverstone_2014}. We also see lower coincidence count rates for cases where we are using TE1 light, due to the higher loss associated with the TE1 light on this chip, as seen in Figure \ref{fig:source_characterisation}. All visibilities in Figure \ref{fig:revhom} are $>90\%$, highlighting how these on-chip sources can effectively generate photon pairs with different TE modes. 

\section{Bell State Generation in Path and TE Mode} \label{sect:BellStateGenSupp}

\begin{figure}
\centering
\includegraphics {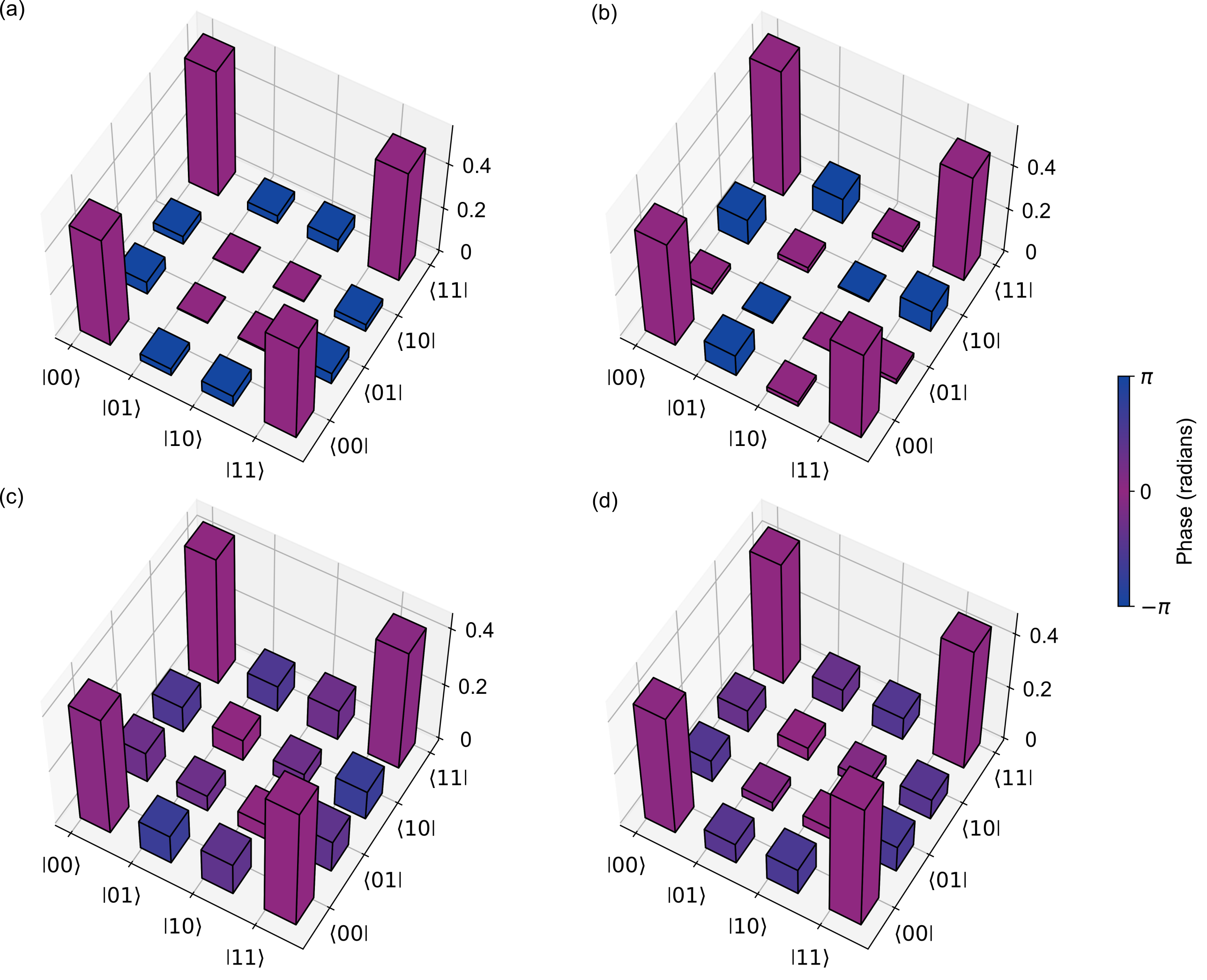}
\caption{\label{fig:bellstates} Reconstructed density matrices for the $\ket{\Phi^+}$ Bell states generated on-chip in the path DOF. (a) Reconstructed state for when TE0 light is sent to both of the spiral sources. (b) Reconstructed state for when TE1 light is sent to both sources. (c) Reconstructed state for sending both TE0 and TE1 light to the bottom spiral source. (d) Reconstructed state for sending both TE0 and TE1 light to the top spiral source.}
\end{figure}

For the DOFs which were used in the final hyperentangled state, we generate and verify each of the constituent Bell states. For the Bell states in path, the initial interferometry prior to the sources was used to pump both the sources with either TE0 or TE1 light. After pair generation, the interferometry after the sources was used to carry out the necessary projective measurements, which were recombined to reconstruct the density matrix using quantum state tomography \cite{Altepeter2004}. To overcome the effects of experimental noise and ensure that the final density matrix reconstruction is physical, maximum likelihood estimation (MLE) is used to find the physical density matrix which most closely corresponds to the outcomes of the projector measurements. The reconstructed Bell states in path are shown in Figure \ref{fig:bellstates}(a,b). The fidelity of this state generated with TE0 light is $\mathcal{F} = 98.9 \pm 0.5 \%$, and the fidelity for the state generated with TE1 light is $ \mathcal{F} = 97.0 \pm 1.6 \%$. Errors are calculated using bootstrapping, as described in Section 5.

A similar method is used to consider the Bell states in mode. The initial interferometry, performed before the sources, is used to pump one of the sources with TE0 and TE1 light. The remaining interferometry is then used to measure the projectors required for tomography, however, due to the absence of phase shifters prior to the MZIs for TE mode operations, an informationally complete set of projectors for the mode qubit cannot be directly measured. Instead, we measure the projectors we can, then perform MLE to establish what the most likely outcome for the other projectors is given our initial data \cite{Hradil2004}, allowing us to reconstruct the full density matrix of the Bell state, similar to the method described in Section 5. The reconstructed Bell states where we consider either sending both modes of light to either the top or bottom source are shown in Figure \ref{fig:bellstates}(c,d). The fidelity for the mode Bell state from the bottom source is $\mathcal{F} = 84.9 \pm 0.2\%$ and from the top source is $\mathcal{F} = 89.9 \pm 0.3\%$.

\section{Techniques for Maximum Likelihood Estimation and Monte Carlo Methods}

 For our measurements of the $\text{GHZ}_4$-style states and the hyperentangled states, additional MLE is utilized. For complete quantum state tomography, the set of projective measurements must satisfy the completeness condition and span the entire operator space. Specifically, for a quantum system of dimension $d$, this requires at least $ d^2 - 1 $ linearly independent measurement outcomes to reconstruct any density matrix uniquely \cite{Sych2012Nov, Rehacek2009Dec}. However, due to the fixed layout of phase shifters on our photonic chip, it is impossible to implement the full set of projectors required for complete quantum state tomography. A full Pauli tomography would involve preparing $ 6^4 = 1296 $ distinct projective measurements for a four-qubit state. Our system, however, is limited to only $4^4=256$ of these configurations. Consequently, the ideal measurement matrix for full tomography should have rank 255, but in our implementation, it only has rank 80. To address this incompleteness, we employ a modified MLE procedure designed for undercomplete measurement settings \cite{Hradil2004Aug}, which still enables the reconstruction of the full estimated density matrix of the hyperentangled state from these incomplete measurements. 

Throughout this work, bootstrapping is used to calculate errors through Monte-Carlo simulations. This involves defining a Poissonian distribution where the mean is the measured probability from the experimental data, and the standard deviation is the error on the measured probability. We sample the probabilities from this distribution, then use these sampled probabilities to calculate the outcome, such as a reconstructed density matrix or fidelity value. The resampling and reconstruction are then repeated until the standard deviation of the metric converges, and the standard error on the mean of the sampled values is taken as our error value. 

\section{Entanglement Entropy}
Entanglement entropy, $\epsilon$, quantifies the degree of entanglement between particles in a quantum state. For a state of two particles, $A,B$, with reduced density matrices $\rho_{A}, \rho_{B}$, we can define the Von Neumann entropy as 
\begin{equation}
    \epsilon = -Tr(\rho_{A} \text{log}_{d} \rho_{A}) = -Tr(\rho_{B} \text{log}_{d} \rho_{B})
\end{equation}
where $d$ is the dimension of the state \cite{nielsen_quantum_2010}. $\epsilon = 0$ denotes a state with no entanglement, and $\epsilon = 1$ denotes a maximally entangled state. 

For our calculations of the entanglement entropy for the GHZ$_{4}$-style state and the hyperentangled state, we quantify the entanglement between the constituent qubits or qudits in the state. To carry out this calculation we initially find the reduced density matrices from the reconstructed density matrix, then use these to calculate the values of entanglement entropy of the state. The average of these entanglement entropy values is then taken.  

\section{Single-Copy Entanglement Distillation}

\begin{figure}
\centering
\includegraphics [scale = 0.875]{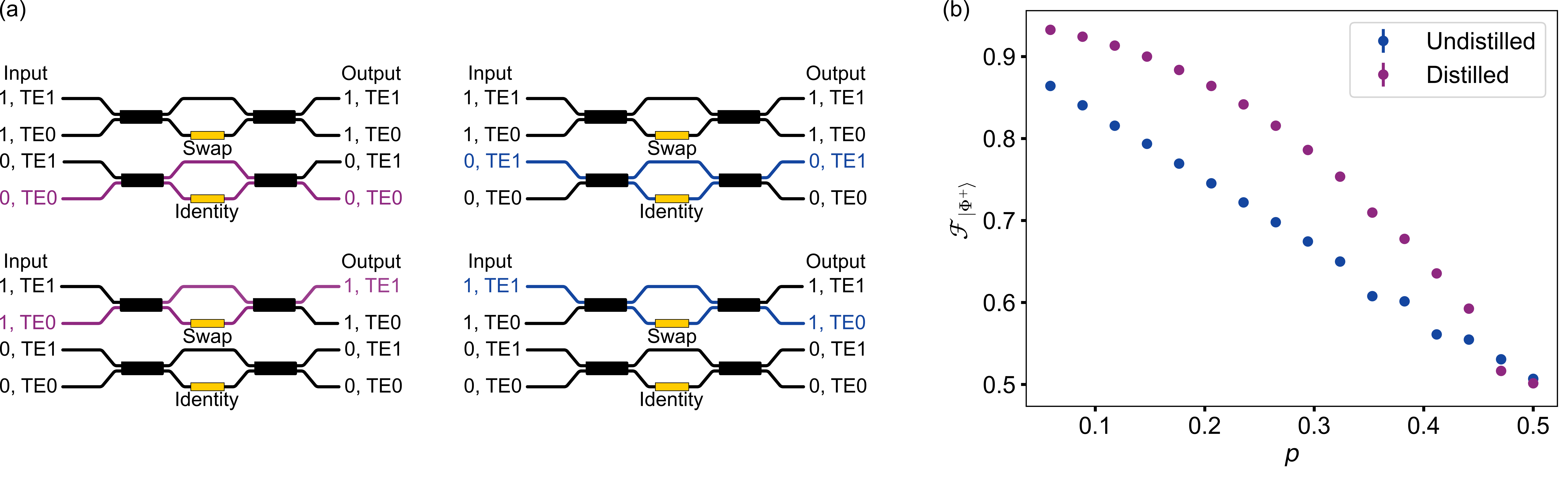}
\caption{\label{fig:cnot}(a) Schematic of how the deterministic CNOT operation for distillation is carried out on the chip using MZIs, with the control being the path DOF and the target being the TE mode. (b) The outcome of the single-copy entanglement distillation experiment between $p = 0$ and $p = 0.5$ where the fidelity is calculated using a reconstructed density matrix from quantum state tomography and MLE.}
\end{figure}

\begin{table}

\newenvironment{tttabular}[1]%
{\begin{tabular}{#1}}%
{\end{tabular}}
\centering
\color{black}
 \begin{tttabular}{|c|c|c|c|}
 \hline
 \multicolumn{2}{|c|}{{Input}}&
 \multicolumn{2}{c|}{{Output}}\\
 \hline
    {Control} & {Target} & {Control} & {Target} \\
     \hline
     \color{mardi_gras} $\Phi^+$ &  \color{mardi_gras} $\Phi^+$ &  \color{mardi_gras}  $\Phi^+$ &  \color{mardi_gras}$\Phi^+$  \\
      \hline
      \color{cobalt_blue}  $\Psi^+$ & \color{mardi_gras} $\Phi^+$ &  \color{cobalt_blue} $\Psi^+$ & \color{cobalt_blue} $\Psi^+$  \\
    \hline
       \color{mardi_gras} $\Phi^+$ &  \color{cobalt_blue}$\Psi^+$ &  \color{mardi_gras} $\Phi^+$ & \color{cobalt_blue}$\Psi^+$  \\
    \hline
        \color{cobalt_blue}$\Psi^+$ &  \color{cobalt_blue}$\Psi^+$ &   \color{cobalt_blue}$\Psi^+$ & 
        \color{mardi_gras}$\Phi^+$  \\
    \hline
 \end{tttabular}
\caption{\label{tab:entdistillbennett} Table adapted from \cite{bennett_purification_1996}, showing the possible states prior to the CNOT for no bit flip, a single bit flip on TE mode or path, bit flips on both mode and path, and the output states following the CNOT operation in the distillation protocol.}
\end{table}

To carry out single-copy entanglement distillation on-chip, bit flip errors need to be set on both the mode and path DOFs, and a deterministic CNOT between the DOFs needs to be carried out on-chip \cite{simon_polarization_2002, ecker_experimental_2021}. For the bit flip errors, these are carried out using MZIs in each DOF, acting on one qubit. For the CNOT, we treat the path as the control and the TE mode as the target, as shown in Figure \ref{fig:cnot}(a). Table \ref{tab:entdistillbennett} shows the input states prior to the distillation protocol in the case of various bit flip errors, and the output states when the CNOT is applied. 

Due to the constraints of the layout of the chip, the CNOT and simulated mode bit flip error are combined on the same MZI in the first section of the interferometer, followed by the simulated path bit flip error. Altering the order of applying these CNOTs and errors within the circuit results in the applied matrix being changed. The matrices for simulating the CNOT with the path error, and for simulating the CNOT with both the path and mode error are swapped, which we account for in our reconstruction of the probabilities. 

To reconstruct the fidelity of the state with varying bit flip error, we initially take data sets for no bit flip error, path bit flip error, mode bit flip error and both path and mode bit flip errors. We then calculate, for some bit flip probability $p$:

\begin{equation}
\begin{split}
P_{\text{none}} &= (1-p)^2, \\
P_{\text{one DOF}} &= p (1-p), \\
P_{\text{both DOFs}} &= p^2, \\
\end{split}
\end{equation}

where the subscript denotes whether or not there is a bit flip, and if it acts on one or both DOFs. We assume that the probability of a bit flip in each DOF is the same. These values are then used to scale our coincidence count data sets:

\begin{equation}
C_{\text{scaled}} = P_{\text{none}}C_{\text{none}} + P_{\text{one DOF}}C_{\text{path}} + P_{\text{one DOF}}C_{\text{mode}} + P_{\text{both DOFs}}C_{\text{mode,path}},
\end{equation}
where $C$ is the normalised coincidence measurements, and the subscript denotes which bit flip errors are applied when this measurement is taken. $C_{scaled}$ is then used to calculate the fidelity from the stabilizers of the path Bell state.

Whilst we use the stabilizer fidelity to consider bit flip errors from $p = 0$ to $p = 1$, we can also use quantum state tomography and MLE to reconstruct the state from $p = 0$ to $p = 0.5$. This is the same method as described in Section \ref{sect:BellStateGenSupp}. Above $p = 0.5$, our target Bell state is no longer dominant and the MLE does not converge. The results for the quantum state tomography and MLE are shown in Figure \ref{fig:cnot}(b). Utilising MLE gives a higher fidelity than calculating from the stabilizers of the state overall, however we still see the same features, such as an improvement at $p = 0$, highlighting how distillation can overcome inherent errors in the system and improve the fidelity. 

\end{document}